\documentclass[aip,graphicx]{revtex4-1}

\usepackage{graphicx}
\usepackage{dcolumn}
\usepackage{bm}
\usepackage[mathlines]{lineno}

\usepackage[utf8]{inputenc}
\usepackage[T1]{fontenc}
\usepackage{mathptmx}
\usepackage{etoolbox}
\usepackage{amssymb}

\makeatletter
\def\@email#1#2{%
 \endgroup
 \patchcmd{\titleblock@produce}
  {\frontmatter@RRAPformat}
  {\frontmatter@RRAPformat{\produce@RRAP{*#1\href{mailto:#2}{#2}}}\frontmatter@RRAPformat}
  {}{}
}%
\makeatother

\draft 

\begin{document}


\title{Amorphous VO$_x$ films with high temperature coefficient of the resistivity grown by reactive e-beam evaporation of V metal} 



\author{E.V. Tarkaeva}
\affiliation{P.N. Lebedev Physical Institute of the Russian Academy of Sciences}
\affiliation{Department of Physics, National Research University Higher School of Economics}

\author{V.A. Ievleva}
\affiliation{Department of Physics, National Research University Higher School of Economics}

\author{A.I. Duleba}
\affiliation{P.N. Lebedev Physical Institute of the Russian Academy of Sciences}

\author{A. V. Muratov}
\affiliation{P.N. Lebedev Physical Institute of the Russian Academy of Sciences}

\author{A.Yu. Kuntsevich}
 \email[Electronic mail: ]{kuncevichay@lebedev.ru}
 \affiliation{P.N. Lebedev Physical Institute of the Russian Academy of Sciences}
 \affiliation{Department of Physics, National Research University Higher School of Economics}


\date{\today}

\begin{abstract}
Amorphous VO$_x$ films without a hysteretic phase transition are stable with respect to thermal cycling and highly demanded as sensitive elements of the resistive thermometers and microbolometers.
In this paper we present simple and low-temperature growth of amorphous vanadium oxide films by reactive electron beam evaporation of vanadium metal in $\sim 10^{-4}$ mBar oxygen atmosphere. The temperature coefficient of the resistivity (TCR) of the films is weakly sensitive to substrate material and temperature and could be tuned by oxygen pressure in the growth chamber up to -2.2\% /K. The resistivity value is stable for months. It depends on the substrate material and substrate temperature during the evaporation.
Simplicity and controllability of the method should lead to various laboratory and industrial applications. 
\end{abstract}


\maketitle 



%
%

%


\section{Introduction}
\label{intro}

Vanadium oxides are fascinating materials with strongly-correlated physics\cite{Kim2004} and yet puzzling phase transitions\cite{Morin1959, Cavalleri(2004)}. On the other hand these materials are stable\cite{stability1}, bio-compatible\cite{bio1} and widely functional\cite{functional}. Strong temperature and deformation dependence of their properties could be used for
bolometric \cite{Wang2013}, thermometric \cite{thermometer1}, thermochromic applications \cite{freestanding}, for strain sensors\cite{strain} and actuators \cite{actuator1}.

Crystalline VO$_2$ demonstrates structural and electronic hysteretic phase transitions at 68 $^\circ$C with the change of crystal lattice symmetry accompanied by strong variation in conductivity and refraction coefficient.
 The transition temperature, hysteresis loop width and the scale of the conductivity change across the transition could be tuned by stoichiometry\cite{stoichiometry}, doping\cite{doping1,doping2} and disorder\cite{disorder}. Importantly, the material preserves strong dependence of the physical properties on temperature even far from the transition.  Phase transition leads to degradation of the films with thermal cycling and to the uncertain temperature dependence of the resistivity.  
From the view of electronic device fabrication it is therefore desirable to  
avoid the transition\cite{epitaxy}.

Amorphous films of VO$_x$, where $x$ is close to 2\cite{amorphous2,amorphous3, amorphous4, amorphous5, amorphous6} are well approved solutions for bolometry and thermometry \cite{thermometer1}. They have strong $T$-dependence of the resistivity without a transition. Typical values of the TCR (temperature coefficient of the resistivity) are about -2\% per K in a wide range of temperatures from 0 to 100 Celsius\cite{thermometer1, amorphous5}).

There are many approaches to the growth of the vanadium oxide thin films, including reactive magnetron sputtering\cite{magnetron1}, epitaxy\cite{epitaxy}, pulsed laser deposition \cite{pld1}, chemical vapour deposition\cite{cvd},  high-temperature oxidation\cite{oxidation} and anodic oxidation\cite{anodicoxidation} of the vanadium films, e-beam evaporation of vanadium oxides\cite{ebeamoxide1,ebeamoxide2,ebeamoxide3}, spray pyrolysis \cite{pyrolysis1, pyrolysis2}, atomic layer deposition\cite{ald}.
 In particular, most of amorphous films reported\cite{amorphous2,amorphous3, amorphous4, amorphous5, amorphous6} are fabricated by reactive magnetron sputtering.
 
As a rule, VO$_x$ thin film fabrication requires rather high temperatures 350-500 $^\circ$C and/or annealing/oxidation at elevated temperatures.
At such temperatures properties of vanadium oxides thin films are extremely sensitive to temperature, exposition time and oxygen concentration\cite{pld1}. Therefore careful adjustment of the film preparation procedure is needed, including  the process gas pressure and composition (O$_2$/Ar ratio). Thus the reactive magnetron growth of amorphous VO$_x$ with high TCR is not a simple process.

 Electron beam evaporation of metals in reactive gas atmospere is an alternative to reactive magnetron sputtering. It was shown to be an effective way to grow amorphous thin films of  SnO$_x$ \cite{Sn}, TiO$_x$ \cite{Ti}, AlO$_x$ and some nitrides \cite{N}. To the best of our knowledge this approach was not previously used  for VO$_x$. 
\begin{figure}[b!]
\centerline{\includegraphics[width=0.5\textwidth]{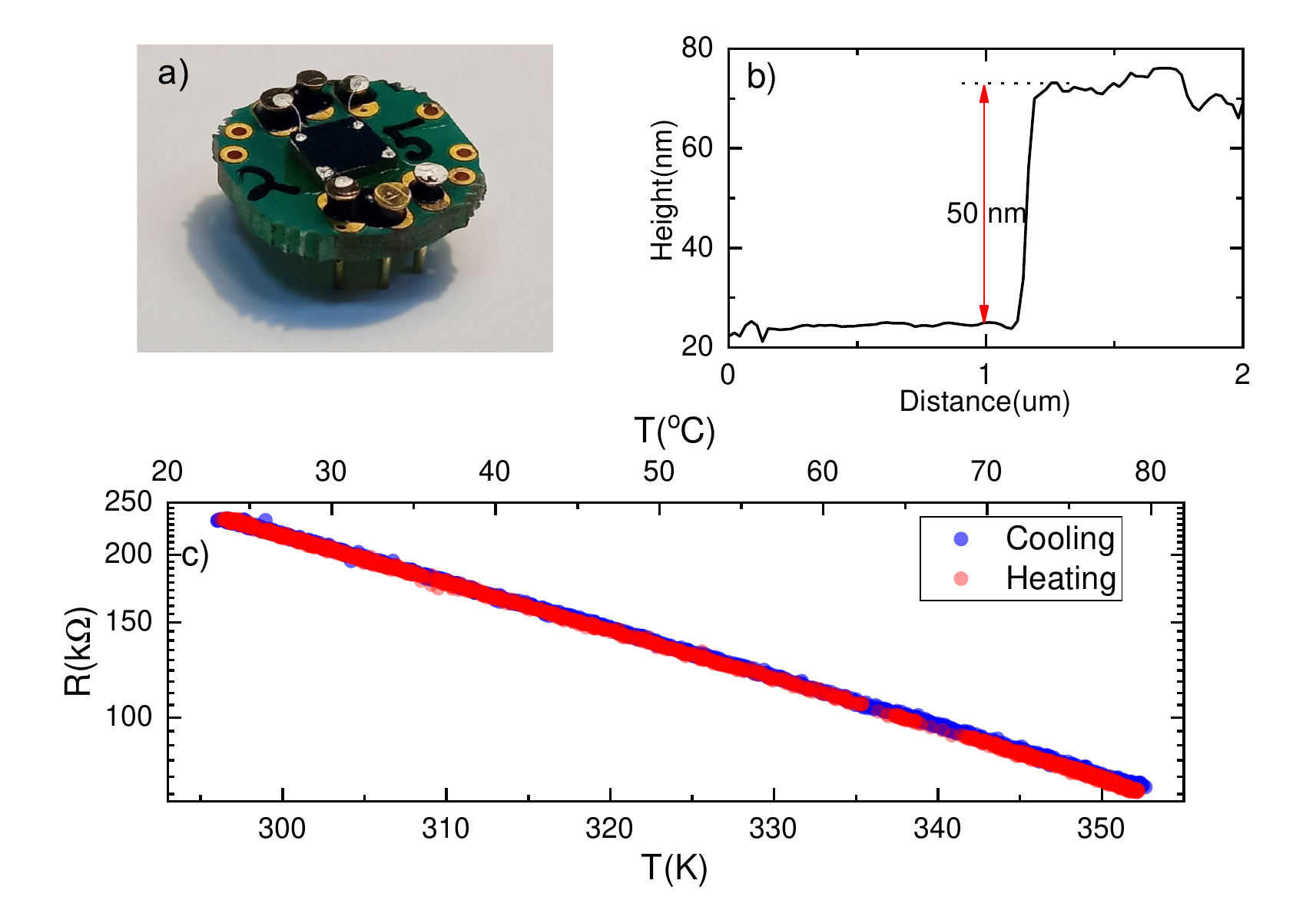}}
\caption{(a) Sample mounted for Van der Pauw transport characterization; (b) AFM profile of the step at the edge of the film; (c) A typical example of the heat-up cooldown $R(T)$ measurement showing no hysteresis.}
\label{Photo_AFM_R(T)hyst}
\end{figure}

In this paper we grow amorphous VO$_x$ films with excellent properties by reactive electron beam evaporation. Vanadium metal target is used for e-beam evaporation in $\sim$10$^{-4}$ mBar  oxygen atmosphere, driven by continious gas flow. This route is low-temperature and very easily reproducible in almost any e-beam evaporation machine. The properties of the obtained films (TCR and resistivity value) are analogous to magnetron films and depend smoothly on three technological parameters: oxygen flow (and hence pressure), evaporation rate and substrate temperature. Optical characterization of the film (Raman, ellipsometry and transmission spectra) confirms the amorphous character and stoichiometry close to VO$_2$. 
We speculate that the evaporated vanadium atoms interact with the oxygen both in the chamber during the flight and at the substrate surface. We also demonstrate compatibility of the film technology with lift-off process thus showing up the application potential. 

\section{Methods}
\label{Methods}

We grow vanadium oxide thin films using Plassys MEB-550S electron beam evaporation machine with multiple crucibles.  Vanadium is evaporated  from tungsten crucible. The distance between the crucible and the substrate is $\approx$0.5 m. The base vacuum of the system is $\sim$3$\cdot 10^{-8}$ mBar. We add 99.9999\% oxygen to the chamber using mass flow controller and measure the pressure in the range from 10$^{-6}$ to $10^{-4}$ mBar.  
For the sake of comparison we evaporate the same film thickness (mostly 30 nm) according to the the quartz monitor calibrated to vanadium.
The substrate temperature could be stabilized in the range 30-700 \textcelsius.

Substrate temperature, oxygen flow and evaporation rate, taken from the quartz thickness monitor are thus the growth technological parameters.
The films are evaporated on different substrates (glass or sapphire). 
The films were characterized using atomic force microscopy in the tapping mode (Solver 47 by NT-MDT).
Temperature dependence of the resistivity of the rectangular pieces in the Van der Pauw geometry was measured using home-made setup on the basis of NI 6351 data acquisition card in the range from 20 to 80 $^\circ$C. A sample mounted for transport measurements is shown in Fig.\ref{Photo_AFM_R(T)hyst}a. Raman spectra at 532 nm were collected using Olympus BX-51 metallographic microscope with EnSpectr R532 express analyzer.

Spectral reflectance and transmittance measurements were performed over the range from 300 to 2500 nm using the J.A. Woollam VASE ellipsometer. 
In order to find refractive and absorption indices from the experimentally measured ellipticity parameters we model  
the sample as consisting of a substrate (glass or sapphire) and thin film and solve numerically  the system of equations:

$$
\Psi(n_f, k_f, d_f, n_s, k_s, \alpha)=\Psi_{exp},
$$$$
\Delta(n_f, k_f, d_f, n_s, k_s, \alpha)=\Delta_{exp}.
$$

Here $\Psi$ and $\Delta$ are ellipticity parameters, $\alpha$ is the angle of incidence, $n_s,\, k_s$ are the indices of refraction and absorption of the substrate, respectively, $d_f$ is the film thickness.

We chose physically plausible solution with continuous smooth spectra that satisfy the requirement $n, k>0$.

\section{Results}
\label{results}
We grew a series of thin films with the same nominal thickness (30nm) and different substrates, growth temperatures, oxygen flows and growth rates. According to AFM Fig.\ref{Photo_AFM_R(T)hyst}b in all VO$_x$ films the real thickness is about 50 nm, that is apparently independent of the technological parameters.  Raman spectra do not show up characteristic peaks. Temperature dependencies of the resistivity has no temperature hysteresis inherent to VO$_2$, as shown in Fig.\ref{Photo_AFM_R(T)hyst}c. All these facts together with ellipsometry measurements (see below) consistently prove the amorphous structure of the films.

\begin{figure}[h]
\centerline{\includegraphics[width=0.8\textwidth]{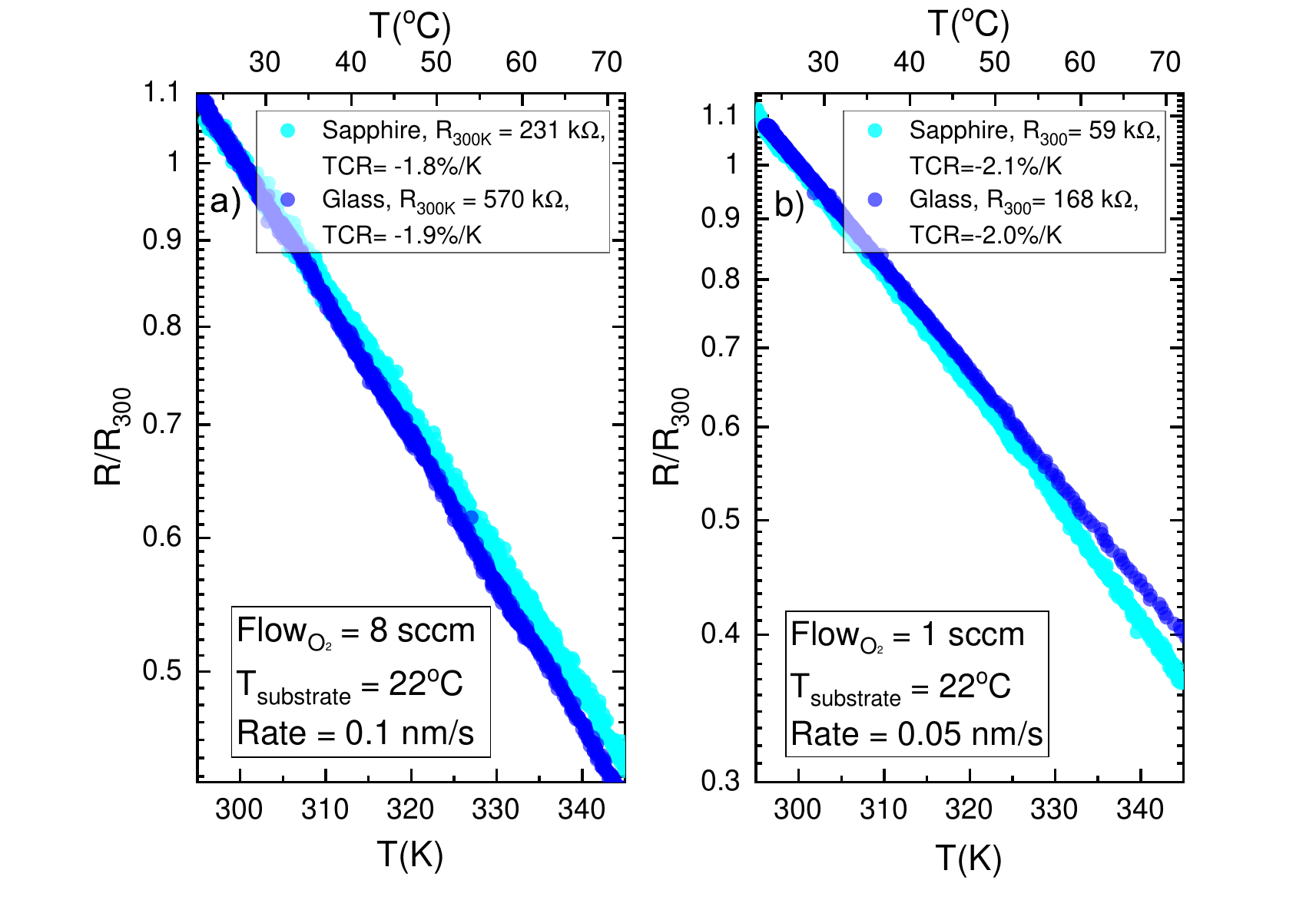}}
\caption{$R(T)$ dependence measured for glass(blue) and sapphire (cyan) substrates in the same growth process. The technological parameters are: a -  oxygen flow - 8 sccm, substrate temperature - 22 \textcelsius, growth rate - 0.1 nm/s; b -  oxygen flow - 1 sccm, substrate temperature - 22 \textcelsius, growth rate - 0.05 nm/s.}

\label{diff_substrate}
\end{figure}
The transport properties depend smoothly on the technological parameters. 
Let us discuss the main results:

\begin{figure}[h]
\centerline{\includegraphics[width=1.1\textwidth]{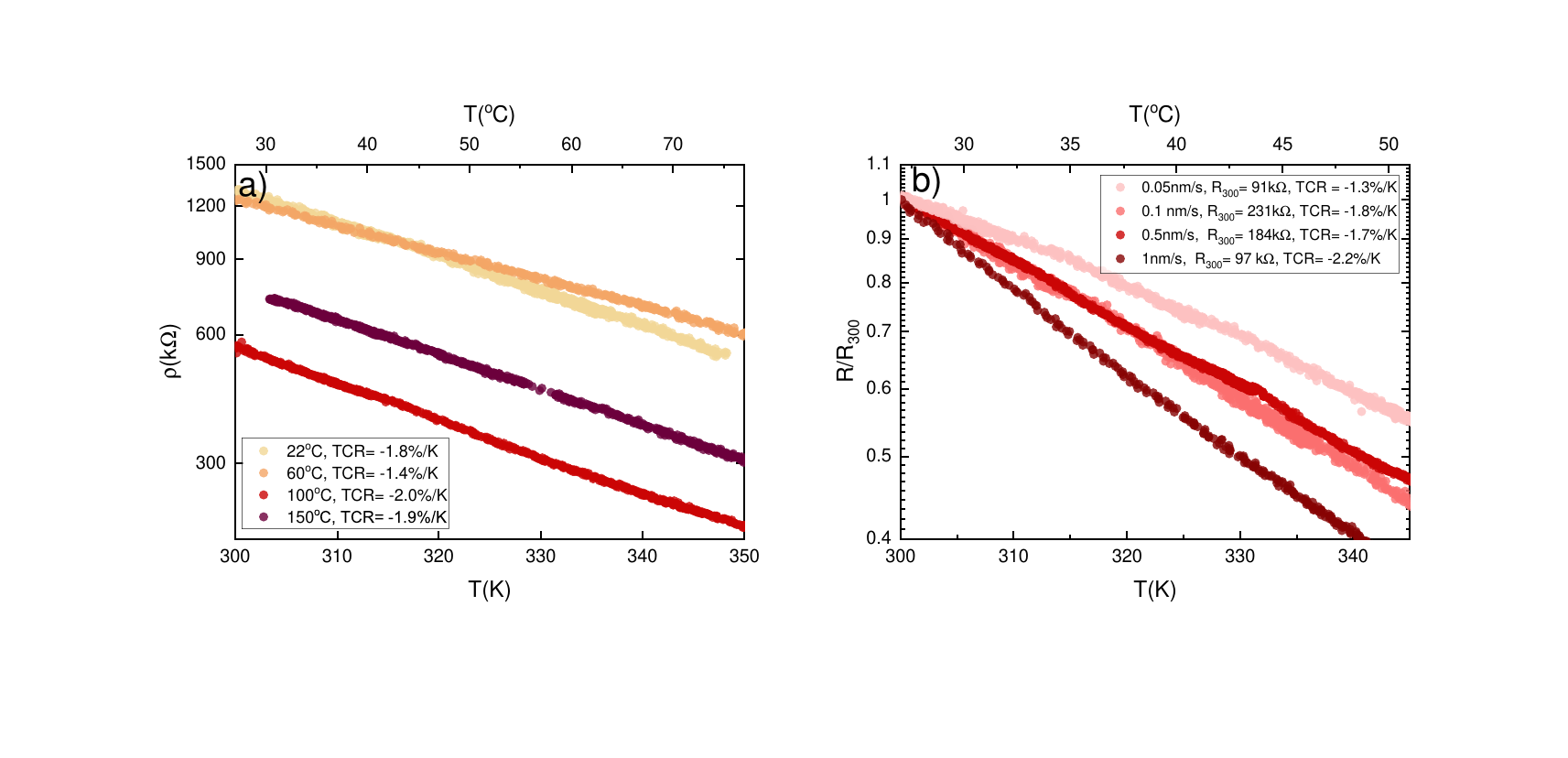}}
\vspace{-0.8in}
\caption{$\rho(T)$ for a series of the films with one technological parameter changing and the other two fixed. a - 
Growth rate - 0.1 nm/s, oxygen flow - 8 sccm. Substrate temperature is changed. b - Substrate temperature - 22 \textcelsius, oxygen flow - 8 sccm, growth rate is changed. }
\label{rate_temp}
\end{figure}

\begin{figure}[h]
\centerline{\includegraphics[width=0.9\textwidth]{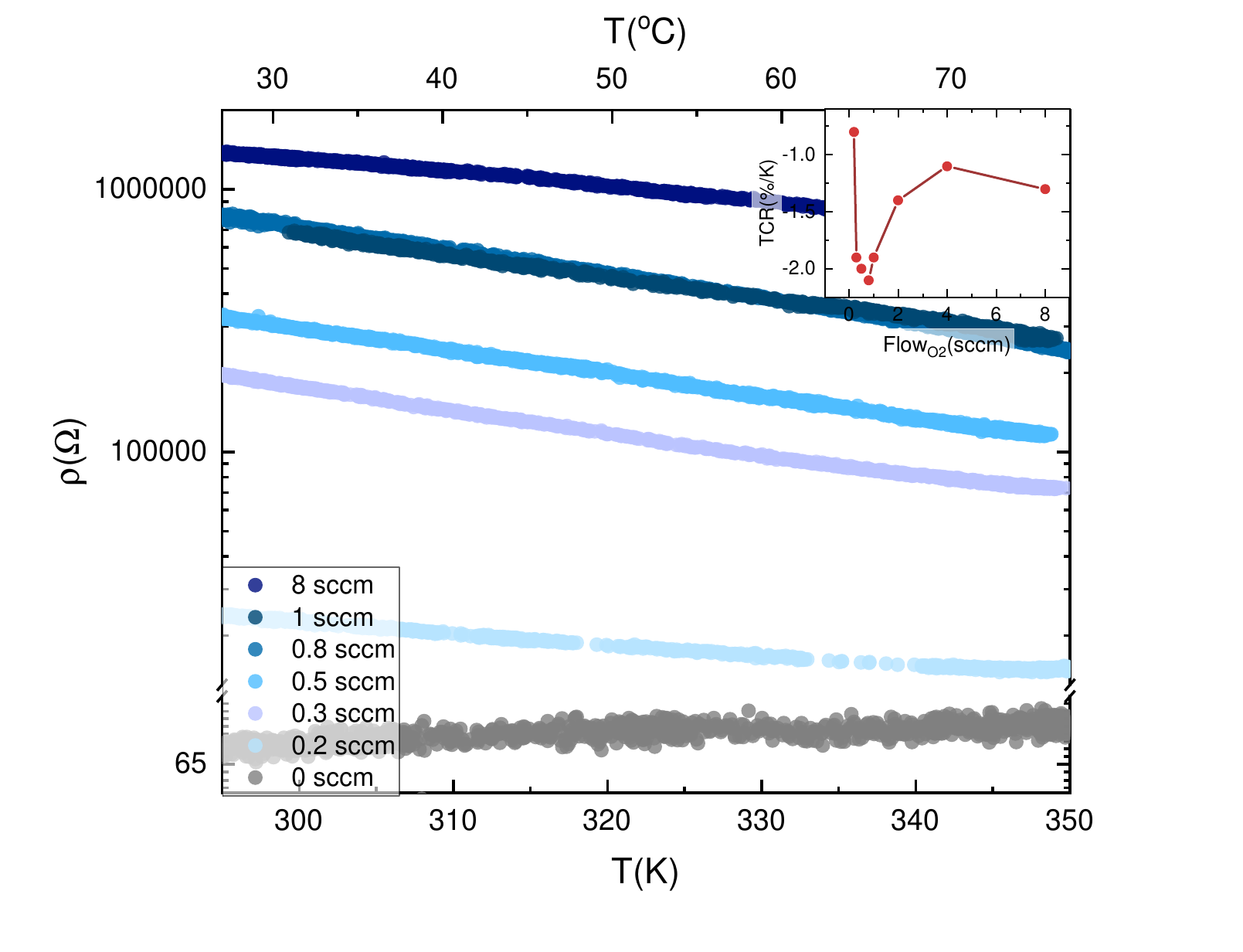}}
\caption{The temperature dependence of the resistivity for different oxygen flow. Insert: the TCR on oxygen flow dependence for all samples with growth rate - 0.05 nm/s and substrate temperature of 22 \textcelsius.}
\label{Sccm}
\end{figure}

\begin{enumerate}
\item{For different substrates (glass and sapphire) the TCR values for the films grown in one process are similar (Fig.\ref{diff_substrate}). At the same time the resistivity value for the film on glass is higher. These results are reproducible for various technological parameters (compare Fig.\ref{diff_substrate}a. and Fig.\ref{diff_substrate}b).} Since the substrate holder is at room temperature and the films are amorphous, the actual temperature at the surface during the growth might be different for glass and sapphire substrates. This difference may affect the size of the VO$_x$ nanograins and the values of the resistivity in turn. 

We present all results below for the sapphire substrate.

\item {For a fixed value of oxygen flow and growth rate the resistivity of the VO$_x$ films does not depend strongly of the substrate temperature, as shown in Fig.\ref{rate_temp}a: the resistivity variation is less than twofold.
TCR does not demonstrate an apparent temperature dependence. 

As a rule, vanadium oxides are grown at much higher substrate temperatures (above 300\textcelsius). Correspondingly, the film properties depend strongly on the substrate temperature.
For example, the only paper on reactive e-beam evaporation of vanadium oxides, that we found \cite{reb}, reports crystalline V$_2$O$_3$ films with properties strongly dependent on substrate temperature in the range 500-800 \textcelsius.

In our work the substrate is relatively cold, therefore the oxidation threshold is mainly covered by the kinetic energy of evaporated vanadium, rather than by thermal motion energy of substrate atoms.

}

\item {For a fixed substrate temperature (22\textcelsius) and oxygen flow (8 sccm) the dependence on growth rate is related to the  stoichiometry of the film (Fig.\ref{rate_temp}b). 

When the growth rate is low, there is excess of oxygen and  the composition is shifted to V$_2$O$_5$: the films have higher resistivity and lower TCR. As the flow increases, the stoichiometry tends to VO$_2$, resistivity decreases and TCR rises \cite{Ionov}. 
Further increase of V flow was technologically not desirable and could damage the crucible.
When we attempted to grow at the maximal rate 1 nm/s and lower oxygen flow (1 sccm), a metallic-like film was formed with several tens Ohm/$\square$ resistivity and almost zero TCR.
}

\item{Dependence of the film resistance on oxygen flow is shown in Fig.\ref{Sccm}. The resistivity at 300K increases smoothly with oxygen flow for a fixed substrate temperature and growth rate. Pure vanadium film with 65 Ohm/\(\square\) resistance  is shown for the reference. An increase in film resistance with an oxygen flow corresponds to the well-known tendency in vanadium oxides \cite{Tendency}.
The non-monotonic dependence of TCR on oxygen flow was observed (see Fig.\ref{Sccm} insert). There is a certain value of oxygen flow (1 sccm) at which TCR reaches it minimum (-2.2 \%/K.), that probably corresponds to optimal stoichiometry.  
This TCR value is in agreement with the results obtained for magnetron-grown amorphous VO$_x$ films \cite{amorphous2,amorphous5}. Similar films are used in commercial bolometric thermal imagers. 
}
\end{enumerate}

\section{Optical measurements (Ellipsometry)}

\begin{figure}[h!]
\centering
\centerline{\includegraphics[width=0.9\textwidth]{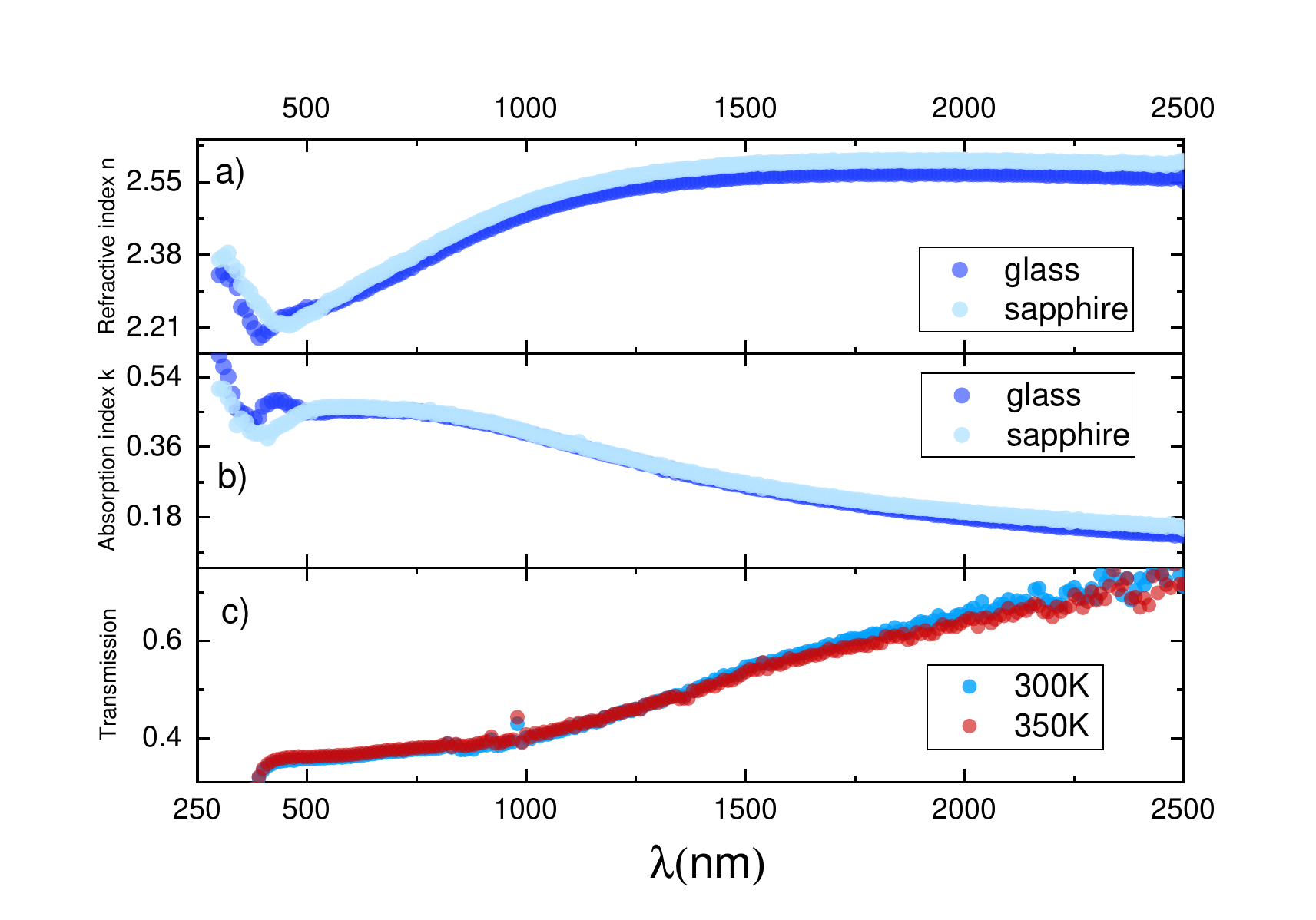}}
\caption{a - wavelength dependence of the refractive index for different substrate material (glass and sapphire), b - wavelength dependence of the absorption index for different substrate material (glass and sapphire), c - transmission spectra of the film on glass substrate at different temperatures.}
\label{Optics}
\end{figure}

For optical measurements we grew two $\approx$80 nm thick films with the optimal growth parameters (1 sccm O$_2$ flow, growth rate 0.05 nm/s and 22\textcelsius \, substrate temperature). 
The optical parameter values appear to be approximately the same on both substrates (Fig.\ref{Optics}a, Fig.\ref{Optics}b), indicating that the substrate material does not affect the film. 

Wavelength dependencies are close to those obtained earlier \cite{VANBILZEN2015143} for amorphous VO$_x$ films. The absorption measurements at different temperatures (300 and 350 K) has shown a minor variation (Fig.\ref{Optics}c), that means absence of crystalline VO$_2$ phase.

\section{Growth mechanism}
\label{mechanism}

\begin{figure}[b!]
\centerline{\includegraphics[scale=0.28]{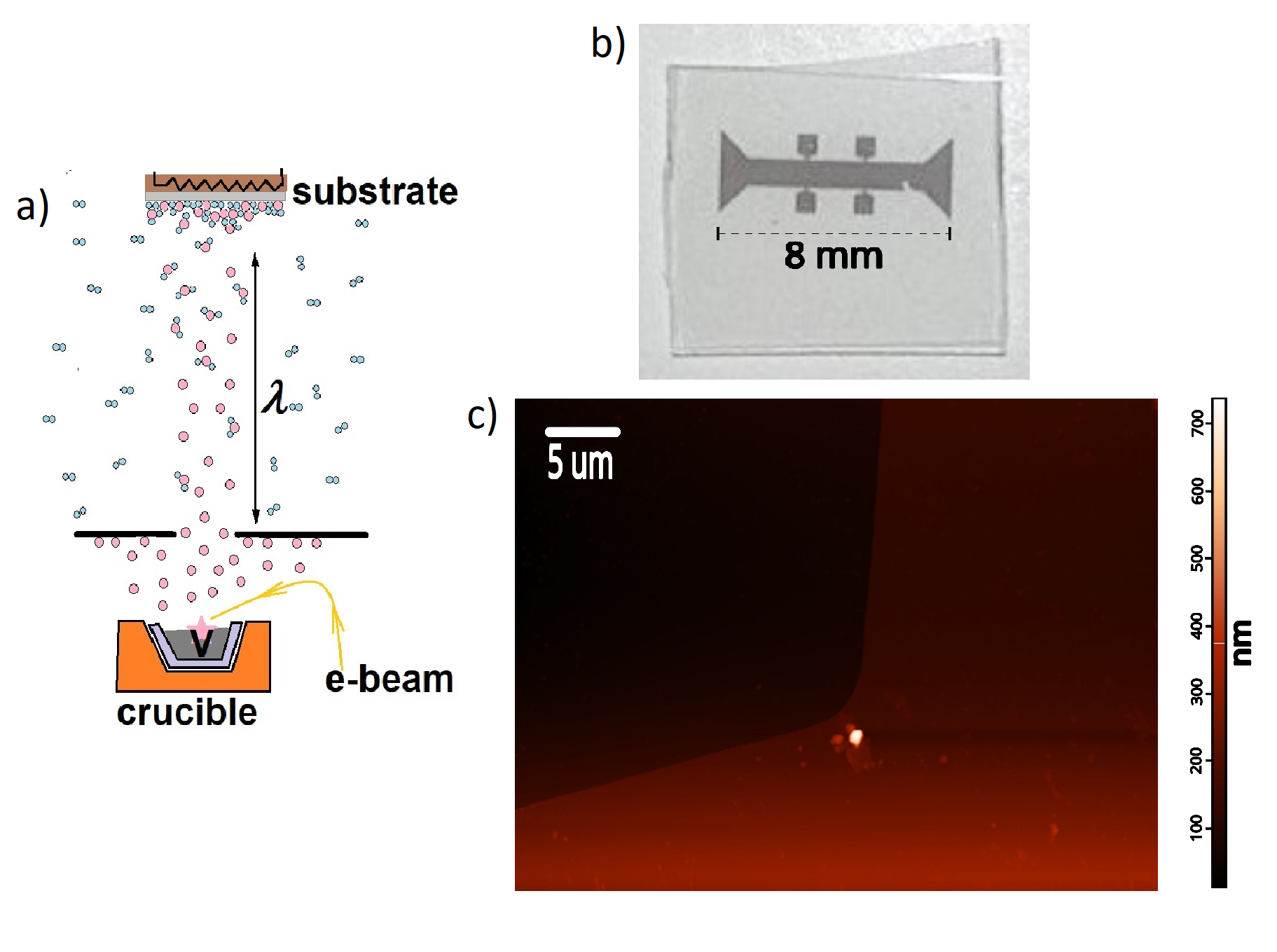}}
\caption{a - process scheme; b - photograph of the lift-off formed mesa-structure on VO$_x$ thin film. c - AFM image of the edge of mesa-structure.}
\label{pS_HB_AFM}
\end{figure}

Vanadium may potentially get oxidized either on the substrate surface or during the flight in the growth chamber. The oxidation of vanadium on a substrate is well known to depend strongly on temperature.  
Independence of the TCR on substrate temperature suggests that essential part of the oxidation occurs on fly, as shown in the scheme of the process (Fig. \ref{pS_HB_AFM}a ).

It is useful to evaluate some figures related to the process. Mass of vanadium atom is 50 a.e.m., while oxygen molecule O$_2$ has a mass of 32 a.e.m. Typical temperature of the evaporated vanadium atoms is about the melting temperature of vanadium 2500 K or even higher.  Oxygen has room temperature $\sim$300 K. Therefore oxygen molecule has almost an order of magnitude smaller momentum than vanadium $p\propto \sqrt{mT}$. Momentum is conserved during the  collision and subsequent chemical reaction. Therefore vanadium does not deflect strongly from its trajectory. 
The growth rate (e.g. 0.5 nm/s ), divided by molecule volume $\sim 5 $ \AA$^{3}$ gives the amount of V atoms per unit area and time, those achieve a substrate $10^{16}$ atoms/(cm$^2$ s). The residual oxygen ($p=2\cdot10^{-4}$ mbar) produces $nV/4$ molecules hits to surface, that is $10^{17}$ atoms/(cm$^2$ s), i.e. the flux of vanadium is smaller than the flux of oxygen. However, the probability of the chemical reaction between the cold surface and oxygen at room temperature is even smaller and the main process is on-fly oxidation. 

Mean free path of vanadium atoms in the residual oxygen gas can be estimated as $1/n \sigma$, where $n$ is the oxygen concentration $p/kT\sim$ 6$\cdot 10^{12}$ atoms/cm$^3$, and  $\sigma$ is vanadium-oxygen scattering cross-section $\pi (r_V+r_{oks})^2\sim 4\cdot10^{-15}$ cm$^{-2}$  (here we use vanadium atomic radius $r_v=1.79$\AA and oxygen molecular radius $r_{oks}=1.52$\AA). Mean free path is about 0.4 m for the typical process gas pressure $p=2\cdot10^{-4}$ mbar. The distance between the electron beam source and the substrate is 0.5 m. In other words, vanadium atom with high probability meets oxygen molecule prior to achievement of a substrate. However for lower pressures mean free path is larger than substrate-to-crucible distance and the oxidation occurs at the substrate. The stoichiometry is shifted towards lower oxygen content as seen from TCR, Fig. \ref{rate_temp}. 

We note that this is an oversimplified qualitative picture. In fact vanadium oxides have a tendency to form clusters up to 10 vanadium atoms\cite{clusters}. Low substrate temperature, arbitrary orientation of the clusters/ molecules and random oxidation of the V atoms does not allow forming crystalline structure of the film.

\section{Discussion. Applications.}

The thickness, TCR and resistivity values of our films are analogous to the reactive magnetron sputtered films from Ref. \cite{amorphous3}. In our case however a much wider range of parameters are tested and smooth systematic dependencies of the properties are demonstrated. 
The main advantage of our growth method is its stability and simplicity. Vanadium metal is much easily available than vanadium oxide targets, there is also not so many requirements to stabilization of the temperature regime or gas mixture composition.
Our process appeared to be so well controllable because the thermal energy of the evaporated vanadium ($\sim 2500$ K) fits well the oxidation reaction, and also because there is a long fly path where interaction with oxygen occurs. In case of magnetron sputtering many technological parameters have to be adjusted to achieve the proper energy of the V atoms and oxide formation. 

The adhesion of the film to the substrate might be improved if the substrate surface is dehydrated e.g. by heating or ion bombardment. Importantly even without this preliminary procedure the adhesion is strong enough so the film survives lift-off process with standard photoresist (AZ 1512 HS). Figure \ref{pS_HB_AFM}b shows a photo of the lift-off formed mesa-stucture, and Fig. \ref{pS_HB_AFM}c shows an AFM image with a smooth edge. Lift-off process allows forming patterns without etching, that might be crucial for device fabrication. For the VO$_x$ grown at elevated temperatures this microelectronic process is impossible because the photoresist will not survive and most probably decompose. High reproducibility of the film properties in combination with lift-off process or stencil lithography\cite{stencil} may allow to pattern resistive thermometers with the pre-defined nominal resistivity on arbitrary substrates. 
Measurement within  Hall bar geometry allowed us check that Van der Pauw value of the resistance in the same film is determined correctly.
Importantly, our films did not show up any signs of degradation after 3 months of storage at ambient conditions.

Further modification of the properties is possible if the films are annealed. For example, In Ref. \cite{amorphous5} similar films were obtained by magnetron sputtering and then annealed.

\section{Conclusions}

We developed a reactive electron beam evaporation method of growth for stable amorphous VO$_x$ thin films, where $x\sim 2$. This multifunctional material is highly demanded for uncooled microbolometers and our films appeared to be similar to those grown by reactive magnetron sputtering. Compared to magnetron sputtering our process requires lower substrate temperature and is better controllable.  We show experimentally and explain smooth dependence of the film properties on technological parameters.




\bibliography{refs}

\begin{thebibliography}{41}%
\makeatletter
\providecommand \@ifxundefined [1]{%
 \@ifx{#1\undefined}
}%
\providecommand \@ifnum [1]{%
 \ifnum #1\expandafter \@firstoftwo
 \else \expandafter \@secondoftwo
 \fi
}%
\providecommand \@ifx [1]{%
 \ifx #1\expandafter \@firstoftwo
 \else \expandafter \@secondoftwo
 \fi
}%
\providecommand \natexlab [1]{#1}%
\providecommand \enquote  [1]{``#1''}%
\providecommand \bibnamefont  [1]{#1}%
\providecommand \bibfnamefont [1]{#1}%
\providecommand \citenamefont [1]{#1}%
\providecommand \href@noop [0]{\@secondoftwo}%
\providecommand \href [0]{\begingroup \@sanitize@url \@href}%
\providecommand \@href[1]{\@@startlink{#1}\@@href}%
\providecommand \@@href[1]{\endgroup#1\@@endlink}%
\providecommand \@sanitize@url [0]{\catcode `\\12\catcode `\$12\catcode
  `\&12\catcode `\#12\catcode `\^12\catcode `\_12\catcode `\%12\relax}%
\providecommand \@@startlink[1]{}%
\providecommand \@@endlink[0]{}%
\providecommand \url  [0]{\begingroup\@sanitize@url \@url }%
\providecommand \@url [1]{\endgroup\@href {#1}{\urlprefix }}%
\providecommand \urlprefix  [0]{URL }%
\providecommand \Eprint [0]{\href }%
\providecommand \doibase [0]{http://dx.doi.org/}%
\providecommand \selectlanguage [0]{\@gobble}%
\providecommand \bibinfo  [0]{\@secondoftwo}%
\providecommand \bibfield  [0]{\@secondoftwo}%
\providecommand \translation [1]{[#1]}%
\providecommand \BibitemOpen [0]{}%
\providecommand \bibitemStop [0]{}%
\providecommand \bibitemNoStop [0]{.\EOS\space}%
\providecommand \EOS [0]{\spacefactor3000\relax}%
\providecommand \BibitemShut  [1]{\csname bibitem#1\endcsname}%
\let\auto@bib@innerbib\@empty
\bibitem [{\citenamefont {Kim}\ \emph {et~al.}(2004)\citenamefont {Kim},
  \citenamefont {Chae}, \citenamefont {Youn}, \citenamefont {Maeng},
  \citenamefont {Kim}, \citenamefont {Kang},\ and\ \citenamefont
  {Lim}}]{Kim2004}%
  \BibitemOpen
  \bibfield  {author} {\bibinfo {author} {\bibfnamefont {H.-T.}\ \bibnamefont
  {Kim}}, \bibinfo {author} {\bibfnamefont {B.-G.}\ \bibnamefont {Chae}},
  \bibinfo {author} {\bibfnamefont {D.-H.}\ \bibnamefont {Youn}}, \bibinfo
  {author} {\bibfnamefont {S.-L.}\ \bibnamefont {Maeng}}, \bibinfo {author}
  {\bibfnamefont {G.}~\bibnamefont {Kim}}, \bibinfo {author} {\bibfnamefont
  {K.-Y.}\ \bibnamefont {Kang}}, \ and\ \bibinfo {author} {\bibfnamefont
  {Y.-S.}\ \bibnamefont {Lim}},\ }\bibfield  {title} {\enquote {\bibinfo
  {title} {Mechanism and observation of mott transition in vo2-based two- and
  three-terminal devices},}\ }\href {\doibase 10.1088/1367-2630/6/1/052}
  {\bibfield  {journal} {\bibinfo  {journal} {New Journal of Physics}\ }\textbf
  {\bibinfo {volume} {6}},\ \bibinfo {pages} {52} (\bibinfo {year}
  {2004})}\BibitemShut {NoStop}%
\bibitem [{\citenamefont {Morin}(1959)}]{Morin1959}%
  \BibitemOpen
  \bibfield  {author} {\bibinfo {author} {\bibfnamefont {F.~J.}\ \bibnamefont
  {Morin}},\ }\bibfield  {title} {\enquote {\bibinfo {title} {Oxides which show
  a metal-to-insulator transition at the neel temperature},}\ }\href {\doibase
  10.1103/PhysRevLett.3.34} {\bibfield  {journal} {\bibinfo  {journal} {Phys.
  Rev. Lett.}\ }\textbf {\bibinfo {volume} {3}},\ \bibinfo {pages} {34--36}
  (\bibinfo {year} {1959})}\BibitemShut {NoStop}%
\bibitem [{\citenamefont {Cavalleri}\ \emph {et~al.}(2004)\citenamefont
  {Cavalleri}, \citenamefont {Chong}, \citenamefont {Fourmaux}, \citenamefont
  {Glover}, \citenamefont {Heimann}, \citenamefont {Kieffer}, \citenamefont
  {Mun}, \citenamefont {Padmore},\ and\ \citenamefont
  {Schoenlein}}]{Cavalleri(2004)}%
  \BibitemOpen
  \bibfield  {author} {\bibinfo {author} {\bibfnamefont {A.}~\bibnamefont
  {Cavalleri}}, \bibinfo {author} {\bibfnamefont {H.~H.~W.}\ \bibnamefont
  {Chong}}, \bibinfo {author} {\bibfnamefont {S.}~\bibnamefont {Fourmaux}},
  \bibinfo {author} {\bibfnamefont {T.~E.}\ \bibnamefont {Glover}}, \bibinfo
  {author} {\bibfnamefont {P.~A.}\ \bibnamefont {Heimann}}, \bibinfo {author}
  {\bibfnamefont {J.~C.}\ \bibnamefont {Kieffer}}, \bibinfo {author}
  {\bibfnamefont {B.~S.}\ \bibnamefont {Mun}}, \bibinfo {author} {\bibfnamefont
  {H.~A.}\ \bibnamefont {Padmore}}, \ and\ \bibinfo {author} {\bibfnamefont
  {R.~W.}\ \bibnamefont {Schoenlein}},\ }\bibfield  {title} {\enquote {\bibinfo
  {title} {Picosecond soft x-ray absorption measurement of the photoinduced
  insulator-to-metal transition in ${\mathrm{vo}}_{2}$},}\ }\href {\doibase
  10.1103/PhysRevB.69.153106} {\bibfield  {journal} {\bibinfo  {journal} {Phys.
  Rev. B}\ }\textbf {\bibinfo {volume} {69}},\ \bibinfo {pages} {153106}
  (\bibinfo {year} {2004})}\BibitemShut {NoStop}%
\bibitem [{\citenamefont {Song}\ \emph {et~al.}(2017)\citenamefont {Song},
  \citenamefont {Liu}, \citenamefont {Yao}, \citenamefont {Kou}, \citenamefont
  {Feng}, \citenamefont {Liu},\ and\ \citenamefont {Li}}]{stability1}%
  \BibitemOpen
  \bibfield  {author} {\bibinfo {author} {\bibfnamefont {Y.}~\bibnamefont
  {Song}}, \bibinfo {author} {\bibfnamefont {T.-Y.}\ \bibnamefont {Liu}},
  \bibinfo {author} {\bibfnamefont {B.}~\bibnamefont {Yao}}, \bibinfo {author}
  {\bibfnamefont {T.-Y.}\ \bibnamefont {Kou}}, \bibinfo {author} {\bibfnamefont
  {D.-Y.}\ \bibnamefont {Feng}}, \bibinfo {author} {\bibfnamefont {X.-X.}\
  \bibnamefont {Liu}}, \ and\ \bibinfo {author} {\bibfnamefont
  {Y.}~\bibnamefont {Li}},\ }\bibfield  {title} {\enquote {\bibinfo {title}
  {Amorphous mixed-valence vanadium oxide/exfoliated carbon cloth structure
  shows a record high cycling stability},}\ }\href {\doibase
  https://doi.org/10.1002/smll.201700067} {\bibfield  {journal} {\bibinfo
  {journal} {Small}\ }\textbf {\bibinfo {volume} {13}},\ \bibinfo {pages}
  {1700067} (\bibinfo {year} {2017})},\ \Eprint
  {http://arxiv.org/abs/https://onlinelibrary.wiley.com/doi/pdf/10.1002/smll.201700067}
  {https://onlinelibrary.wiley.com/doi/pdf/10.1002/smll.201700067} \BibitemShut
  {NoStop}%
\bibitem [{\citenamefont {Nie}\ \emph {et~al.}(2014)\citenamefont {Nie},
  \citenamefont {Zhang}, \citenamefont {Lei}, \citenamefont {Yang},
  \citenamefont {Zhang}, \citenamefont {Lu},\ and\ \citenamefont
  {Wang}}]{bio1}%
  \BibitemOpen
  \bibfield  {author} {\bibinfo {author} {\bibfnamefont {G.}~\bibnamefont
  {Nie}}, \bibinfo {author} {\bibfnamefont {L.}~\bibnamefont {Zhang}}, \bibinfo
  {author} {\bibfnamefont {J.}~\bibnamefont {Lei}}, \bibinfo {author}
  {\bibfnamefont {L.}~\bibnamefont {Yang}}, \bibinfo {author} {\bibfnamefont
  {Z.}~\bibnamefont {Zhang}}, \bibinfo {author} {\bibfnamefont
  {X.}~\bibnamefont {Lu}}, \ and\ \bibinfo {author} {\bibfnamefont
  {C.}~\bibnamefont {Wang}},\ }\bibfield  {title} {\enquote {\bibinfo {title}
  {Monocrystalline vo2 (b) nanobelts: large-scale synthesis{,} intrinsic
  peroxidase-like activity and application in biosensing},}\ }\href {\doibase
  10.1039/C3TA15051H} {\bibfield  {journal} {\bibinfo  {journal} {J. Mater.
  Chem. A}\ }\textbf {\bibinfo {volume} {2}},\ \bibinfo {pages} {2910--2914}
  (\bibinfo {year} {2014})}\BibitemShut {NoStop}%
\bibitem [{\citenamefont {Hu}\ \emph {et~al.}(2023)\citenamefont {Hu},
  \citenamefont {Hu}, \citenamefont {Vu}, \citenamefont {Li}, \citenamefont
  {Wang}, \citenamefont {Ke}, \citenamefont {Zeng}, \citenamefont {Mai},\ and\
  \citenamefont {Long}}]{functional}%
  \BibitemOpen
  \bibfield  {author} {\bibinfo {author} {\bibfnamefont {P.}~\bibnamefont
  {Hu}}, \bibinfo {author} {\bibfnamefont {P.}~\bibnamefont {Hu}}, \bibinfo
  {author} {\bibfnamefont {T.~D.}\ \bibnamefont {Vu}}, \bibinfo {author}
  {\bibfnamefont {M.}~\bibnamefont {Li}}, \bibinfo {author} {\bibfnamefont
  {S.}~\bibnamefont {Wang}}, \bibinfo {author} {\bibfnamefont {Y.}~\bibnamefont
  {Ke}}, \bibinfo {author} {\bibfnamefont {X.}~\bibnamefont {Zeng}}, \bibinfo
  {author} {\bibfnamefont {L.}~\bibnamefont {Mai}}, \ and\ \bibinfo {author}
  {\bibfnamefont {Y.}~\bibnamefont {Long}},\ }\bibfield  {title} {\enquote
  {\bibinfo {title} {Vanadium oxide: Phase diagrams, structures, synthesis, and
  applications},}\ }\href {\doibase 10.1021/acs.chemrev.2c00546} {\bibfield
  {journal} {\bibinfo  {journal} {Chemical Reviews}\ }\textbf {\bibinfo
  {volume} {123}},\ \bibinfo {pages} {4353--4415} (\bibinfo {year}
  {2023})}\BibitemShut {NoStop}%
\bibitem [{\citenamefont {Wang}\ \emph {et~al.}(2013)\citenamefont {Wang},
  \citenamefont {Lai}, \citenamefont {Li}, \citenamefont {Hu},\ and\
  \citenamefont {Chen}}]{Wang2013}%
  \BibitemOpen
  \bibfield  {author} {\bibinfo {author} {\bibfnamefont {B.}~\bibnamefont
  {Wang}}, \bibinfo {author} {\bibfnamefont {J.}~\bibnamefont {Lai}}, \bibinfo
  {author} {\bibfnamefont {H.}~\bibnamefont {Li}}, \bibinfo {author}
  {\bibfnamefont {H.}~\bibnamefont {Hu}}, \ and\ \bibinfo {author}
  {\bibfnamefont {S.}~\bibnamefont {Chen}},\ }\bibfield  {title} {\enquote
  {\bibinfo {title} {Nanostructured vanadium oxide thin film with high tcr at
  room temperature for microbolometer},}\ }\href {\doibase
  10.1016/j.infrared.2012.10.006} {\bibfield  {journal} {\bibinfo  {journal}
  {Infrared Physics and Technology}\ }\textbf {\bibinfo {volume} {57}},\
  \bibinfo {pages} {8–13} (\bibinfo {year} {2013})}\BibitemShut {NoStop}%
\bibitem [{\citenamefont {Zia}\ \emph {et~al.}(2017)\citenamefont {Zia},
  \citenamefont {Abdel-Rahman}, \citenamefont {Alduraibi}, \citenamefont
  {Ilahi}, \citenamefont {Awad},\ and\ \citenamefont {Majzoub}}]{thermometer1}%
  \BibitemOpen
  \bibfield  {author} {\bibinfo {author} {\bibfnamefont {M.~F.}\ \bibnamefont
  {Zia}}, \bibinfo {author} {\bibfnamefont {M.}~\bibnamefont {Abdel-Rahman}},
  \bibinfo {author} {\bibfnamefont {M.}~\bibnamefont {Alduraibi}}, \bibinfo
  {author} {\bibfnamefont {B.}~\bibnamefont {Ilahi}}, \bibinfo {author}
  {\bibfnamefont {E.}~\bibnamefont {Awad}}, \ and\ \bibinfo {author}
  {\bibfnamefont {S.}~\bibnamefont {Majzoub}},\ }\bibfield  {title} {\enquote
  {\bibinfo {title} {Electrical and infrared optical properties of vanadium
  oxide semiconducting thin-film thermometers},}\ }\href {\doibase
  10.1007/s11664-017-5571-0} {\bibfield  {journal} {\bibinfo  {journal}
  {Journal of Electronic Materials}\ }\textbf {\bibinfo {volume} {46}},\
  \bibinfo {pages} {5978--5985} (\bibinfo {year} {2017})}\BibitemShut {NoStop}%
\bibitem [{\citenamefont {Ma}\ \emph {et~al.}(2021)\citenamefont {Ma},
  \citenamefont {Xiao}, \citenamefont {Wang}, \citenamefont {Sun},
  \citenamefont {Wang}, \citenamefont {Gao}, \citenamefont {Wang},
  \citenamefont {Jiang}, \citenamefont {Liu},\ and\ \citenamefont
  {Zhang}}]{freestanding}%
  \BibitemOpen
  \bibfield  {author} {\bibinfo {author} {\bibfnamefont {H.}~\bibnamefont
  {Ma}}, \bibinfo {author} {\bibfnamefont {X.}~\bibnamefont {Xiao}}, \bibinfo
  {author} {\bibfnamefont {Y.}~\bibnamefont {Wang}}, \bibinfo {author}
  {\bibfnamefont {Y.}~\bibnamefont {Sun}}, \bibinfo {author} {\bibfnamefont
  {B.}~\bibnamefont {Wang}}, \bibinfo {author} {\bibfnamefont {X.}~\bibnamefont
  {Gao}}, \bibinfo {author} {\bibfnamefont {E.}~\bibnamefont {Wang}}, \bibinfo
  {author} {\bibfnamefont {K.}~\bibnamefont {Jiang}}, \bibinfo {author}
  {\bibfnamefont {K.}~\bibnamefont {Liu}}, \ and\ \bibinfo {author}
  {\bibfnamefont {X.}~\bibnamefont {Zhang}},\ }\bibfield  {title} {\enquote
  {\bibinfo {title} {Wafer-scale freestanding vanadium dioxide film},}\ }\href
  {\doibase 10.1126/sciadv.abk3438} {\bibfield  {journal} {\bibinfo  {journal}
  {Science Advances}\ }\textbf {\bibinfo {volume} {7}},\ \bibinfo {pages}
  {eabk3438} (\bibinfo {year} {2021})},\ \Eprint
  {http://arxiv.org/abs/https://www.science.org/doi/pdf/10.1126/sciadv.abk3438}
  {https://www.science.org/doi/pdf/10.1126/sciadv.abk3438} \BibitemShut
  {NoStop}%
\bibitem [{\citenamefont {Hu}\ \emph {et~al.}(2010)\citenamefont {Hu},
  \citenamefont {Ding}, \citenamefont {Chen}, \citenamefont {Kulkarni},
  \citenamefont {Shen}, \citenamefont {Tsukruk},\ and\ \citenamefont
  {Wang}}]{strain}%
  \BibitemOpen
  \bibfield  {author} {\bibinfo {author} {\bibfnamefont {B.}~\bibnamefont
  {Hu}}, \bibinfo {author} {\bibfnamefont {Y.}~\bibnamefont {Ding}}, \bibinfo
  {author} {\bibfnamefont {W.}~\bibnamefont {Chen}}, \bibinfo {author}
  {\bibfnamefont {D.}~\bibnamefont {Kulkarni}}, \bibinfo {author}
  {\bibfnamefont {Y.}~\bibnamefont {Shen}}, \bibinfo {author} {\bibfnamefont
  {V.}~\bibnamefont {Tsukruk}}, \ and\ \bibinfo {author} {\bibfnamefont
  {Z.}~\bibnamefont {Wang}},\ }\bibfield  {title} {\enquote {\bibinfo {title}
  {External-strain induced insulating phase transition in vo2 nanobeam and its
  application as flexible strain sensor},}\ }\href {\doibase
  10.1002/adma.201002868} {\bibfield  {journal} {\bibinfo  {journal} {Advanced
  materials (Deerfield Beach, Fla.)}\ }\textbf {\bibinfo {volume} {22}},\
  \bibinfo {pages} {5134--9} (\bibinfo {year} {2010})}\BibitemShut {NoStop}%
\bibitem [{\citenamefont {Liu}\ \emph {et~al.}(2014)\citenamefont {Liu},
  \citenamefont {C}, \citenamefont {Suh}, \citenamefont {Tang-Kong},
  \citenamefont {Fu}, \citenamefont {Lee}, \citenamefont {Zhou}, \citenamefont
  {Chua},\ and\ \citenamefont {Wu}}]{actuator1}%
  \BibitemOpen
  \bibfield  {author} {\bibinfo {author} {\bibfnamefont {K.}~\bibnamefont
  {Liu}}, \bibinfo {author} {\bibfnamefont {C.}~\bibnamefont {C}}, \bibinfo
  {author} {\bibfnamefont {J.}~\bibnamefont {Suh}}, \bibinfo {author}
  {\bibfnamefont {R.}~\bibnamefont {Tang-Kong}}, \bibinfo {author}
  {\bibfnamefont {D.}~\bibnamefont {Fu}}, \bibinfo {author} {\bibfnamefont
  {S.}~\bibnamefont {Lee}}, \bibinfo {author} {\bibfnamefont {J.}~\bibnamefont
  {Zhou}}, \bibinfo {author} {\bibfnamefont {L.}~\bibnamefont {Chua}}, \ and\
  \bibinfo {author} {\bibfnamefont {J.}~\bibnamefont {Wu}},\ }\bibfield
  {title} {\enquote {\bibinfo {title} {Powerful, multifunctional torsional
  micromuscles activated by phase transition},}\ }\href {\doibase
  10.1002/adma.201304064} {\bibfield  {journal} {\bibinfo  {journal} {Advanced
  materials (Deerfield Beach, Fla.)}\ }\textbf {\bibinfo {volume} {26}},\
  \bibinfo {pages} {1746--50} (\bibinfo {year} {2014})}\BibitemShut {NoStop}%
\bibitem [{\citenamefont {Griffiths}\ and\ \citenamefont
  {Eastwood}(2003)}]{stoichiometry}%
  \BibitemOpen
  \bibfield  {author} {\bibinfo {author} {\bibfnamefont {C.~H.}\ \bibnamefont
  {Griffiths}}\ and\ \bibinfo {author} {\bibfnamefont {H.~K.}\ \bibnamefont
  {Eastwood}},\ }\bibfield  {title} {\enquote {\bibinfo {title} {{Influence of
  stoichiometry on the metal‐semiconductor transition in vanadium
  dioxide}},}\ }\href {\doibase 10.1063/1.1663568} {\bibfield  {journal}
  {\bibinfo  {journal} {Journal of Applied Physics}\ }\textbf {\bibinfo
  {volume} {45}},\ \bibinfo {pages} {2201--2206} (\bibinfo {year} {2003})},\
  \Eprint
  {http://arxiv.org/abs/https://pubs.aip.org/aip/jap/article-pdf/45/5/2201/10566252/2201\_1\_online.pdf}
  {https://pubs.aip.org/aip/jap/article-pdf/45/5/2201/10566252/2201\_1\_online.pdf}
  \BibitemShut {NoStop}%
\bibitem [{\citenamefont {Soltani}\ \emph {et~al.}(2004)\citenamefont
  {Soltani}, \citenamefont {Chaker}, \citenamefont {Haddad}, \citenamefont
  {Kruzelecky},\ and\ \citenamefont {Margot}}]{doping1}%
  \BibitemOpen
  \bibfield  {author} {\bibinfo {author} {\bibfnamefont {M.}~\bibnamefont
  {Soltani}}, \bibinfo {author} {\bibfnamefont {M.}~\bibnamefont {Chaker}},
  \bibinfo {author} {\bibfnamefont {E.}~\bibnamefont {Haddad}}, \bibinfo
  {author} {\bibfnamefont {R.~V.}\ \bibnamefont {Kruzelecky}}, \ and\ \bibinfo
  {author} {\bibfnamefont {J.}~\bibnamefont {Margot}},\ }\bibfield  {title}
  {\enquote {\bibinfo {title} {{Effects of Ti–W codoping on the optical and
  electrical switching of vanadium dioxide thin films grown by a reactive
  pulsed laser deposition}},}\ }\href {\doibase 10.1063/1.1788883} {\bibfield
  {journal} {\bibinfo  {journal} {Applied Physics Letters}\ }\textbf {\bibinfo
  {volume} {85}},\ \bibinfo {pages} {1958--1960} (\bibinfo {year} {2004})},\
  \Eprint
  {http://arxiv.org/abs/https://pubs.aip.org/aip/apl/article-pdf/85/11/1958/13164512/1958\_1\_online.pdf}
  {https://pubs.aip.org/aip/apl/article-pdf/85/11/1958/13164512/1958\_1\_online.pdf}
  \BibitemShut {NoStop}%
\bibitem [{\citenamefont {Émond}, \citenamefont {Hendaoui},\ and\
  \citenamefont {Chaker}(2015)}]{doping2}%
  \BibitemOpen
  \bibfield  {author} {\bibinfo {author} {\bibfnamefont {N.}~\bibnamefont
  {Émond}}, \bibinfo {author} {\bibfnamefont {A.}~\bibnamefont {Hendaoui}}, \
  and\ \bibinfo {author} {\bibfnamefont {M.}~\bibnamefont {Chaker}},\
  }\bibfield  {title} {\enquote {\bibinfo {title} {Low resistivity
  w$_x$v$_{1-x}$o$_2$-based multilayer structure with high temperature
  coefficient of resistance for microbolometer applications},}\ }\href
  {\doibase 10.1063/1.4932954} {\bibfield  {journal} {\bibinfo  {journal}
  {Applied Physics Letters}\ }\textbf {\bibinfo {volume} {107}},\ \bibinfo
  {pages} {143507} (\bibinfo {year} {2015})}\BibitemShut {NoStop}%
\bibitem [{\citenamefont {Ramirez}\ \emph {et~al.}(2015)\citenamefont
  {Ramirez}, \citenamefont {Saerbeck}, \citenamefont {Wang}, \citenamefont
  {Trastoy}, \citenamefont {Malnou}, \citenamefont {Lesueur}, \citenamefont
  {Crocombette}, \citenamefont {Villegas},\ and\ \citenamefont
  {Schuller}}]{disorder}%
  \BibitemOpen
  \bibfield  {author} {\bibinfo {author} {\bibfnamefont {J.~G.}\ \bibnamefont
  {Ramirez}}, \bibinfo {author} {\bibfnamefont {T.}~\bibnamefont {Saerbeck}},
  \bibinfo {author} {\bibfnamefont {S.}~\bibnamefont {Wang}}, \bibinfo {author}
  {\bibfnamefont {J.}~\bibnamefont {Trastoy}}, \bibinfo {author} {\bibfnamefont
  {M.}~\bibnamefont {Malnou}}, \bibinfo {author} {\bibfnamefont
  {J.}~\bibnamefont {Lesueur}}, \bibinfo {author} {\bibfnamefont {J.-P.}\
  \bibnamefont {Crocombette}}, \bibinfo {author} {\bibfnamefont {J.~E.}\
  \bibnamefont {Villegas}}, \ and\ \bibinfo {author} {\bibfnamefont {I.~K.}\
  \bibnamefont {Schuller}},\ }\bibfield  {title} {\enquote {\bibinfo {title}
  {Effect of disorder on the metal-insulator transition of vanadium oxides:
  Local versus global effects},}\ }\href {\doibase 10.1103/PhysRevB.91.205123}
  {\bibfield  {journal} {\bibinfo  {journal} {Phys. Rev. B}\ }\textbf {\bibinfo
  {volume} {91}},\ \bibinfo {pages} {205123} (\bibinfo {year}
  {2015})}\BibitemShut {NoStop}%
\bibitem [{\citenamefont {Ainabayev}\ \emph {et~al.}(2023)\citenamefont
  {Ainabayev}, \citenamefont {Mullarkey}, \citenamefont {Walls}, \citenamefont
  {Caffrey}, \citenamefont {Zhussupbekov}, \citenamefont {Zhussupbekova},
  \citenamefont {Ilhan}, \citenamefont {Kaisha}, \citenamefont {Biswas},
  \citenamefont {Tikhonov}, \citenamefont {Murtagh},\ and\ \citenamefont
  {Shvets}}]{epitaxy}%
  \BibitemOpen
  \bibfield  {author} {\bibinfo {author} {\bibfnamefont {A.}~\bibnamefont
  {Ainabayev}}, \bibinfo {author} {\bibfnamefont {D.}~\bibnamefont
  {Mullarkey}}, \bibinfo {author} {\bibfnamefont {B.}~\bibnamefont {Walls}},
  \bibinfo {author} {\bibfnamefont {D.}~\bibnamefont {Caffrey}}, \bibinfo
  {author} {\bibfnamefont {K.}~\bibnamefont {Zhussupbekov}}, \bibinfo {author}
  {\bibfnamefont {A.}~\bibnamefont {Zhussupbekova}}, \bibinfo {author}
  {\bibfnamefont {C.}~\bibnamefont {Ilhan}}, \bibinfo {author} {\bibfnamefont
  {A.}~\bibnamefont {Kaisha}}, \bibinfo {author} {\bibfnamefont
  {P.}~\bibnamefont {Biswas}}, \bibinfo {author} {\bibfnamefont
  {A.}~\bibnamefont {Tikhonov}}, \bibinfo {author} {\bibfnamefont
  {O.}~\bibnamefont {Murtagh}}, \ and\ \bibinfo {author} {\bibfnamefont
  {I.}~\bibnamefont {Shvets}},\ }\bibfield  {title} {\enquote {\bibinfo {title}
  {Epitaxial grown vo2 with suppressed hysteresis and low room temperature
  resistivity for high-performance thermal sensor applications},}\ }\href
  {\doibase 10.1021/acsanm.2c05297} {\bibfield  {journal} {\bibinfo  {journal}
  {ACS Applied Nano Materials}\ }\textbf {\bibinfo {volume} {6}},\ \bibinfo
  {pages} {2917--2927} (\bibinfo {year} {2023})}\BibitemShut {NoStop}%
\bibitem [{\citenamefont {Rehman}\ and\ \citenamefont
  {Dezhi}(2015)}]{amorphous2}%
  \BibitemOpen
  \bibfield  {author} {\bibinfo {author} {\bibfnamefont {F.}~\bibnamefont
  {Rehman}}\ and\ \bibinfo {author} {\bibfnamefont {S.}~\bibnamefont {Dezhi}},\
  }\bibfield  {title} {\enquote {\bibinfo {title} {Evolution of microstructure
  in vanadium oxide bolometer ﬁlm during annealing process},}\ }\href
  {\doibase 10.1016/j.apsusc.2015.09.068} {\bibfield  {journal} {\bibinfo
  {journal} {Applied Surface Science}\ }\textbf {\bibinfo {volume} {357}},\
  \bibinfo {pages} {887–891} (\bibinfo {year} {2015})}\BibitemShut {NoStop}%
\bibitem [{\citenamefont {Chen}, \citenamefont {Jiang},\ and\ \citenamefont
  {Li}(2014)}]{amorphous3}%
  \BibitemOpen
  \bibfield  {author} {\bibinfo {author} {\bibfnamefont {R.-H.}\ \bibnamefont
  {Chen}}, \bibinfo {author} {\bibfnamefont {Y.-L.}\ \bibnamefont {Jiang}}, \
  and\ \bibinfo {author} {\bibfnamefont {B.}~\bibnamefont {Li}},\ }\bibfield
  {title} {\enquote {\bibinfo {title} {Influence of post-annealing on
  resistivity of vox thin film},}\ }\href
  {https://api.semanticscholar.org/CorpusID:6260491} {\bibfield  {journal}
  {\bibinfo  {journal} {IEEE Electron Device Letters}\ }\textbf {\bibinfo
  {volume} {35}},\ \bibinfo {pages} {780--782} (\bibinfo {year}
  {2014})}\BibitemShut {NoStop}%
\bibitem [{\citenamefont {Podraza}\ \emph {et~al.}(2012)\citenamefont
  {Podraza}, \citenamefont {Gauntt}, \citenamefont {Motyka}, \citenamefont
  {Dickey},\ and\ \citenamefont {Horn}}]{amorphous4}%
  \BibitemOpen
  \bibfield  {author} {\bibinfo {author} {\bibfnamefont {N.~J.}\ \bibnamefont
  {Podraza}}, \bibinfo {author} {\bibfnamefont {B.~D.}\ \bibnamefont {Gauntt}},
  \bibinfo {author} {\bibfnamefont {M.~A.}\ \bibnamefont {Motyka}}, \bibinfo
  {author} {\bibfnamefont {E.~C.}\ \bibnamefont {Dickey}}, \ and\ \bibinfo
  {author} {\bibfnamefont {M.~W.}\ \bibnamefont {Horn}},\ }\bibfield  {title}
  {\enquote {\bibinfo {title} {{Electrical and optical properties of sputtered
  amorphous vanadium oxide thin films}},}\ }\href {\doibase 10.1063/1.3702451}
  {\bibfield  {journal} {\bibinfo  {journal} {Journal of Applied Physics}\
  }\textbf {\bibinfo {volume} {111}},\ \bibinfo {pages} {073522} (\bibinfo
  {year} {2012})},\ \Eprint
  {http://arxiv.org/abs/https://pubs.aip.org/aip/jap/article-pdf/doi/10.1063/1.3702451/15085237/073522\_1\_online.pdf}
  {https://pubs.aip.org/aip/jap/article-pdf/doi/10.1063/1.3702451/15085237/073522\_1\_online.pdf}
  \BibitemShut {NoStop}%
\bibitem [{\citenamefont {Venkatasubramanian}\ \emph
  {et~al.}(2009)\citenamefont {Venkatasubramanian}, \citenamefont {Cabarcos},
  \citenamefont {Allara}, \citenamefont {Horn},\ and\ \citenamefont
  {Ashok}}]{amorphous5}%
  \BibitemOpen
  \bibfield  {author} {\bibinfo {author} {\bibfnamefont {C.}~\bibnamefont
  {Venkatasubramanian}}, \bibinfo {author} {\bibfnamefont {O.~M.}\ \bibnamefont
  {Cabarcos}}, \bibinfo {author} {\bibfnamefont {D.~L.}\ \bibnamefont
  {Allara}}, \bibinfo {author} {\bibfnamefont {M.~W.}\ \bibnamefont {Horn}}, \
  and\ \bibinfo {author} {\bibfnamefont {S.}~\bibnamefont {Ashok}},\ }\bibfield
   {title} {\enquote {\bibinfo {title} {{Correlation of temperature response
  and structure of annealed VOx thin films for IR detector applications}},}\
  }\href {\doibase 10.1116/1.3143667} {\bibfield  {journal} {\bibinfo
  {journal} {Journal of Vacuum Science and Technology A}\ }\textbf {\bibinfo
  {volume} {27}},\ \bibinfo {pages} {956--961} (\bibinfo {year} {2009})},\
  \Eprint
  {http://arxiv.org/abs/https://pubs.aip.org/avs/jva/article-pdf/27/4/956/13302659/956\_1\_online.pdf}
  {https://pubs.aip.org/avs/jva/article-pdf/27/4/956/13302659/956\_1\_online.pdf}
  \BibitemShut {NoStop}%
\bibitem [{\citenamefont {Petnikota}\ \emph {et~al.}(2018)\citenamefont
  {Petnikota}, \citenamefont {Chua}, \citenamefont {Zhou}, \citenamefont
  {Edison},\ and\ \citenamefont {Srinivasan}}]{amorphous6}%
  \BibitemOpen
  \bibfield  {author} {\bibinfo {author} {\bibfnamefont {S.}~\bibnamefont
  {Petnikota}}, \bibinfo {author} {\bibfnamefont {R.}~\bibnamefont {Chua}},
  \bibinfo {author} {\bibfnamefont {Y.}~\bibnamefont {Zhou}}, \bibinfo {author}
  {\bibfnamefont {E.}~\bibnamefont {Edison}}, \ and\ \bibinfo {author}
  {\bibfnamefont {M.}~\bibnamefont {Srinivasan}},\ }\bibfield  {title}
  {\enquote {\bibinfo {title} {Amorphous vanadium oxide thin films as stable
  performing cathodes of lithium and sodium-ion batteries},}\ }\href {\doibase
  10.1186/s11671-018-2766-0} {\bibfield  {journal} {\bibinfo  {journal}
  {Nanoscale Research Letters}\ }\textbf {\bibinfo {volume} {13}},\ \bibinfo
  {pages} {363} (\bibinfo {year} {2018})}\BibitemShut {NoStop}%
\bibitem [{\citenamefont {Zhu}\ \emph {et~al.}(2021)\citenamefont {Zhu},
  \citenamefont {Zhang}, \citenamefont {Jiang}, \citenamefont {Liu},
  \citenamefont {Qi},\ and\ \citenamefont {Yang}}]{magnetron1}%
  \BibitemOpen
  \bibfield  {author} {\bibinfo {author} {\bibfnamefont {M.}~\bibnamefont
  {Zhu}}, \bibinfo {author} {\bibfnamefont {D.}~\bibnamefont {Zhang}}, \bibinfo
  {author} {\bibfnamefont {S.}~\bibnamefont {Jiang}}, \bibinfo {author}
  {\bibfnamefont {S.}~\bibnamefont {Liu}}, \bibinfo {author} {\bibfnamefont
  {H.}~\bibnamefont {Qi}}, \ and\ \bibinfo {author} {\bibfnamefont
  {Y.}~\bibnamefont {Yang}},\ }\bibfield  {title} {\enquote {\bibinfo {title}
  {Phase evolution and thermochromism of vanadium oxide thin films grown at low
  substrate temperatures during magnetron sputtering},}\ }\href {\doibase
  10.1016/j.ceramint.2021.02.115} {\bibfield  {journal} {\bibinfo  {journal}
  {Ceramics International}\ }\textbf {\bibinfo {volume} {47}} (\bibinfo {year}
  {2021}),\ 10.1016/j.ceramint.2021.02.115}\BibitemShut {NoStop}%
\bibitem [{\citenamefont {{Bukhari}}\ \emph {et~al.}(2020)\citenamefont
  {{Bukhari}}, \citenamefont {{Kumar}}, \citenamefont {{Kumar}}, \citenamefont
  {{Gumfekar}}, \citenamefont {{Chung}}, \citenamefont {{Thundat}},\ and\
  \citenamefont {{Goswami}}}]{pld1}%
  \BibitemOpen
  \bibfield  {author} {\bibinfo {author} {\bibfnamefont {S.~A.}\ \bibnamefont
  {{Bukhari}}}, \bibinfo {author} {\bibfnamefont {S.}~\bibnamefont {{Kumar}}},
  \bibinfo {author} {\bibfnamefont {P.}~\bibnamefont {{Kumar}}}, \bibinfo
  {author} {\bibfnamefont {S.~P.}\ \bibnamefont {{Gumfekar}}}, \bibinfo
  {author} {\bibfnamefont {H.-J.}\ \bibnamefont {{Chung}}}, \bibinfo {author}
  {\bibfnamefont {T.}~\bibnamefont {{Thundat}}}, \ and\ \bibinfo {author}
  {\bibfnamefont {A.}~\bibnamefont {{Goswami}}},\ }\bibfield  {title} {\enquote
  {\bibinfo {title} {{The effect of oxygen flow rate on metal-insulator
  transition (MIT) characteristics of vanadium dioxide (VO$_{2}$) thin films by
  pulsed laser deposition (PLD)}},}\ }\href {\doibase
  10.1016/j.apsusc.2020.146995} {\bibfield  {journal} {\bibinfo  {journal}
  {Applied Surface Science}\ }\textbf {\bibinfo {volume} {529}},\ \bibinfo
  {eid} {146995} (\bibinfo {year} {2020})}\BibitemShut {NoStop}%
\bibitem [{\citenamefont {Bahlawane}\ and\ \citenamefont
  {Lenoble}(2014)}]{cvd}%
  \BibitemOpen
  \bibfield  {author} {\bibinfo {author} {\bibfnamefont {N.}~\bibnamefont
  {Bahlawane}}\ and\ \bibinfo {author} {\bibfnamefont {D.}~\bibnamefont
  {Lenoble}},\ }\bibfield  {title} {\enquote {\bibinfo {title} {Vanadium oxide
  compounds: Structure, properties, and growth from the gas phase},}\ }\href
  {\doibase 10.1002/cvde.201400057} {\bibfield  {journal} {\bibinfo  {journal}
  {Chemical Vapor Deposition}\ }\textbf {\bibinfo {volume} {20}} (\bibinfo
  {year} {2014}),\ 10.1002/cvde.201400057}\BibitemShut {NoStop}%
\bibitem [{\citenamefont {Pedrosa}\ \emph {et~al.}(2016)\citenamefont
  {Pedrosa}, \citenamefont {Martin}, \citenamefont {Salut}, \citenamefont {Arab
  Pour~Yazdi},\ and\ \citenamefont {Billard}}]{oxidation}%
  \BibitemOpen
  \bibfield  {author} {\bibinfo {author} {\bibfnamefont {P.}~\bibnamefont
  {Pedrosa}}, \bibinfo {author} {\bibfnamefont {N.}~\bibnamefont {Martin}},
  \bibinfo {author} {\bibfnamefont {R.}~\bibnamefont {Salut}}, \bibinfo
  {author} {\bibfnamefont {M.}~\bibnamefont {Arab Pour~Yazdi}}, \ and\ \bibinfo
  {author} {\bibfnamefont {A.}~\bibnamefont {Billard}},\ }\bibfield  {title}
  {\enquote {\bibinfo {title} {Controlled thermal oxidation of nanostructured
  vanadium thin films},}\ }\href {\doibase 10.1016/j.matlet.2016.03.097}
  {\bibfield  {journal} {\bibinfo  {journal} {Materials Letters}\ }\textbf
  {\bibinfo {volume} {174}} (\bibinfo {year} {2016}),\
  10.1016/j.matlet.2016.03.097}\BibitemShut {NoStop}%
\bibitem [{\citenamefont {Stefanovich}\ \emph {et~al.}(2004)\citenamefont
  {Stefanovich}, \citenamefont {Pergament}, \citenamefont {Velichko},\ and\
  \citenamefont {Stefanovich}}]{anodicoxidation}%
  \BibitemOpen
  \bibfield  {author} {\bibinfo {author} {\bibfnamefont {G.}~\bibnamefont
  {Stefanovich}}, \bibinfo {author} {\bibfnamefont {A.}~\bibnamefont
  {Pergament}}, \bibinfo {author} {\bibfnamefont {A.}~\bibnamefont {Velichko}},
  \ and\ \bibinfo {author} {\bibfnamefont {L.}~\bibnamefont {Stefanovich}},\
  }\bibfield  {title} {\enquote {\bibinfo {title} {Anodic oxidation of vanadium
  and properties of vanadium oxide films},}\ }\href {\doibase
  10.1088/0953-8984/16/23/018} {\bibfield  {journal} {\bibinfo  {journal}
  {Journal of Physics: Condensed Matter}\ }\textbf {\bibinfo {volume} {16}},\
  \bibinfo {pages} {4013} (\bibinfo {year} {2004})}\BibitemShut {NoStop}%
\bibitem [{\citenamefont {Aryasomayajula}, \citenamefont {Reddy},\ and\
  \citenamefont {Nagendra}(2008)}]{ebeamoxide1}%
  \BibitemOpen
  \bibfield  {author} {\bibinfo {author} {\bibfnamefont {S.}~\bibnamefont
  {Aryasomayajula}}, \bibinfo {author} {\bibfnamefont {Y.}~\bibnamefont
  {Reddy}}, \ and\ \bibinfo {author} {\bibfnamefont {C.}~\bibnamefont
  {Nagendra}},\ }\bibfield  {title} {\enquote {\bibinfo {title} {Nano-vanadium
  oxide thin films in mixed phase for microbolometer applications},}\ }\href
  {\doibase 10.1088/0022-3727/41/19/195108} {\bibfield  {journal} {\bibinfo
  {journal} {Journal of Physics D: Applied Physics}\ }\textbf {\bibinfo
  {volume} {41}},\ \bibinfo {pages} {195108} (\bibinfo {year}
  {2008})}\BibitemShut {NoStop}%
\bibitem [{\citenamefont {Marvel}\ \emph {et~al.}(2012)\citenamefont {Marvel},
  \citenamefont {Appavoo}, \citenamefont {Choi}, \citenamefont {Nag},\ and\
  \citenamefont {Haglund}}]{ebeamoxide2}%
  \BibitemOpen
  \bibfield  {author} {\bibinfo {author} {\bibfnamefont {R.}~\bibnamefont
  {Marvel}}, \bibinfo {author} {\bibfnamefont {K.}~\bibnamefont {Appavoo}},
  \bibinfo {author} {\bibfnamefont {B.}~\bibnamefont {Choi}}, \bibinfo {author}
  {\bibfnamefont {J.}~\bibnamefont {Nag}}, \ and\ \bibinfo {author}
  {\bibfnamefont {R.}~\bibnamefont {Haglund}},\ }\bibfield  {title} {\enquote
  {\bibinfo {title} {Electron-beam deposition of vanadium dioxide thin
  films},}\ }\href {\doibase 10.1007/s00339-012-7324-5} {\bibfield  {journal}
  {\bibinfo  {journal} {Applied Physics A}\ ,\ \bibinfo {pages} {1--7}}
  (\bibinfo {year} {2012})}\BibitemShut {NoStop}%
\bibitem [{\citenamefont {{Ramana}}\ \emph {et~al.}(1998)\citenamefont
  {{Ramana}}, \citenamefont {{Hussain}}, \citenamefont {{Uthanna}},\ and\
  \citenamefont {{Srinivasulu Naidu}}}]{ebeamoxide3}%
  \BibitemOpen
  \bibfield  {author} {\bibinfo {author} {\bibfnamefont {C.~V.}\ \bibnamefont
  {{Ramana}}}, \bibinfo {author} {\bibfnamefont {O.~M.}\ \bibnamefont
  {{Hussain}}}, \bibinfo {author} {\bibfnamefont {S.}~\bibnamefont
  {{Uthanna}}}, \ and\ \bibinfo {author} {\bibfnamefont {B.}~\bibnamefont
  {{Srinivasulu Naidu}}},\ }\bibfield  {title} {\enquote {\bibinfo {title}
  {{Influence of oxygen partial pressure on the optical properties of electron
  beam evaporated vanadium pentoxide thin films}},}\ }\href {\doibase
  10.1016/S0925-3467(97)00168-7} {\bibfield  {journal} {\bibinfo  {journal}
  {Optical Materials}\ }\textbf {\bibinfo {volume} {10}},\ \bibinfo {pages}
  {101--107} (\bibinfo {year} {1998})}\BibitemShut {NoStop}%
\bibitem [{\citenamefont {Benkahoul}\ \emph {et~al.}(2017)\citenamefont
  {Benkahoul}, \citenamefont {Zayed}, \citenamefont {Solieman},\ and\
  \citenamefont {Alamri}}]{pyrolysis1}%
  \BibitemOpen
  \bibfield  {author} {\bibinfo {author} {\bibfnamefont {M.}~\bibnamefont
  {Benkahoul}}, \bibinfo {author} {\bibfnamefont {M.}~\bibnamefont {Zayed}},
  \bibinfo {author} {\bibfnamefont {A.}~\bibnamefont {Solieman}}, \ and\
  \bibinfo {author} {\bibfnamefont {S.~N.}\ \bibnamefont {Alamri}},\ }\bibfield
   {title} {\enquote {\bibinfo {title} {Spray deposition of v4o9 and v2o5 thin
  films and post-annealing formation of thermochromic vo2},}\ }\href {\doibase
  10.1016/j.jallcom.2017.02.088} {\bibfield  {journal} {\bibinfo  {journal}
  {Journal of Alloys and Compounds}\ }\textbf {\bibinfo {volume} {704}}
  (\bibinfo {year} {2017}),\ 10.1016/j.jallcom.2017.02.088}\BibitemShut
  {NoStop}%
\bibitem [{\citenamefont {Tadeo}\ \emph {et~al.}(2019)\citenamefont {Tadeo},
  \citenamefont {Panzi}, \citenamefont {Krupanidhi},\ and\ \citenamefont
  {Umarji}}]{pyrolysis2}%
  \BibitemOpen
  \bibfield  {author} {\bibinfo {author} {\bibfnamefont {I.}~\bibnamefont
  {Tadeo}}, \bibinfo {author} {\bibfnamefont {M.}~\bibnamefont {Panzi}},
  \bibinfo {author} {\bibfnamefont {S.}~\bibnamefont {Krupanidhi}}, \ and\
  \bibinfo {author} {\bibfnamefont {A.}~\bibnamefont {Umarji}},\ }\bibfield
  {title} {\enquote {\bibinfo {title} {Low-cost vo$_2$ (m1) thin films
  synthesized by ultrasonic nebulized spray pyrolysis of an aqueous combustion
  mixture for ir photodetection},}\ }\href {\doibase 10.1039/C9RA00189A}
  {\bibfield  {journal} {\bibinfo  {journal} {RSC Advances}\ }\textbf {\bibinfo
  {volume} {9}},\ \bibinfo {pages} {9983--9992} (\bibinfo {year}
  {2019})}\BibitemShut {NoStop}%
\bibitem [{\citenamefont {Blanquart}\ \emph {et~al.}(2013)\citenamefont
  {Blanquart}, \citenamefont {Niinistö}, \citenamefont {Gavagnin},
  \citenamefont {Longo}, \citenamefont {Heikkilä}, \citenamefont
  {Puukilainen}, \citenamefont {Pallem}, \citenamefont {Dussarrat},
  \citenamefont {Ritala},\ and\ \citenamefont {Leskelä}}]{ald}%
  \BibitemOpen
  \bibfield  {author} {\bibinfo {author} {\bibfnamefont {T.}~\bibnamefont
  {Blanquart}}, \bibinfo {author} {\bibfnamefont {J.}~\bibnamefont
  {Niinistö}}, \bibinfo {author} {\bibfnamefont {M.}~\bibnamefont {Gavagnin}},
  \bibinfo {author} {\bibfnamefont {V.}~\bibnamefont {Longo}}, \bibinfo
  {author} {\bibfnamefont {M.}~\bibnamefont {Heikkilä}}, \bibinfo {author}
  {\bibfnamefont {E.}~\bibnamefont {Puukilainen}}, \bibinfo {author}
  {\bibfnamefont {V.}~\bibnamefont {Pallem}}, \bibinfo {author} {\bibfnamefont
  {C.}~\bibnamefont {Dussarrat}}, \bibinfo {author} {\bibfnamefont
  {M.}~\bibnamefont {Ritala}}, \ and\ \bibinfo {author} {\bibfnamefont
  {M.}~\bibnamefont {Leskelä}},\ }\bibfield  {title} {\enquote {\bibinfo
  {title} {Atomic layer deposition and characterization of vanadium oxide thin
  films},}\ }\href {\doibase 10.1039/C2RA22820C} {\bibfield  {journal}
  {\bibinfo  {journal} {RSC Advances}\ }\textbf {\bibinfo {volume} {3}},\
  \bibinfo {pages} {1179--1185} (\bibinfo {year} {2013})}\BibitemShut {NoStop}%
\bibitem [{\citenamefont {Behrouznejad}(2023)}]{Sn}%
  \BibitemOpen
  \bibfield  {author} {\bibinfo {author} {\bibfnamefont {F.}~\bibnamefont
  {Behrouznejad}},\ }\bibfield  {title} {\enquote {\bibinfo {title} {Reactive
  e-beam evaporated snox layer as an effective etl for highly efficient
  spray-coated perovskite solar cells},}\ }\href {\doibase
  https://doi.org/10.1016/j.matchemphys.2023.128086} {\bibfield  {journal}
  {\bibinfo  {journal} {Materials Chemistry and Physics}\ }\textbf {\bibinfo
  {volume} {306}},\ \bibinfo {pages} {128086} (\bibinfo {year}
  {2023})}\BibitemShut {NoStop}%
\bibitem [{\citenamefont {Lin}\ \emph {et~al.}(2009)\citenamefont {Lin},
  \citenamefont {Lee}, \citenamefont {Choi}, \citenamefont {Noh},\ and\
  \citenamefont {Chung}}]{Ti}%
  \BibitemOpen
  \bibfield  {author} {\bibinfo {author} {\bibfnamefont {Z.}~\bibnamefont
  {Lin}}, \bibinfo {author} {\bibfnamefont {I.-S.}\ \bibnamefont {Lee}},
  \bibinfo {author} {\bibfnamefont {Y.-J.}\ \bibnamefont {Choi}}, \bibinfo
  {author} {\bibfnamefont {I.-S.}\ \bibnamefont {Noh}}, \ and\ \bibinfo
  {author} {\bibfnamefont {S.-M.}\ \bibnamefont {Chung}},\ }\bibfield  {title}
  {\enquote {\bibinfo {title} {Characterizations of the tio2-x films
  synthesized by e-beam evaporation for endovascular applications},}\ }\href
  {\doibase 10.1088/1748-6041/4/1/015013} {\bibfield  {journal} {\bibinfo
  {journal} {Biomedical materials (Bristol, England)}\ }\textbf {\bibinfo
  {volume} {4}},\ \bibinfo {pages} {015013} (\bibinfo {year}
  {2009})}\BibitemShut {NoStop}%
\bibitem [{\citenamefont {Mencia}, \citenamefont {Lin},\ and\ \citenamefont
  {Manucharyan}(2021)}]{N}%
  \BibitemOpen
  \bibfield  {author} {\bibinfo {author} {\bibfnamefont {R.}~\bibnamefont
  {Mencia}}, \bibinfo {author} {\bibfnamefont {Y.-H.}\ \bibnamefont {Lin}}, \
  and\ \bibinfo {author} {\bibfnamefont {V.}~\bibnamefont {Manucharyan}},\
  }\bibfield  {title} {\enquote {\bibinfo {title} {{Superconducting titanium
  nitride films grown by directional reactive evaporation}},}\ }\href {\doibase
  10.1063/5.0048819} {\bibfield  {journal} {\bibinfo  {journal} {Journal of
  Applied Physics}\ }\textbf {\bibinfo {volume} {130}},\ \bibinfo {pages}
  {225109} (\bibinfo {year} {2021})},\ \Eprint
  {http://arxiv.org/abs/https://pubs.aip.org/aip/jap/article-pdf/doi/10.1063/5.0048819/15273893/225109\_1\_online.pdf}
  {https://pubs.aip.org/aip/jap/article-pdf/doi/10.1063/5.0048819/15273893/225109\_1\_online.pdf}
  \BibitemShut {NoStop}%
\bibitem [{\citenamefont {Schuler}\ \emph {et~al.}(1995)\citenamefont
  {Schuler}, \citenamefont {Weissmann}, \citenamefont {Renner}, \citenamefont
  {Six}, \citenamefont {Klimm}, \citenamefont {Simmet},\ and\ \citenamefont
  {Horn}}]{reb}%
  \BibitemOpen
  \bibfield  {author} {\bibinfo {author} {\bibfnamefont {H.}~\bibnamefont
  {Schuler}}, \bibinfo {author} {\bibfnamefont {G.}~\bibnamefont {Weissmann}},
  \bibinfo {author} {\bibfnamefont {C.}~\bibnamefont {Renner}}, \bibinfo
  {author} {\bibfnamefont {S.}~\bibnamefont {Six}}, \bibinfo {author}
  {\bibfnamefont {S.}~\bibnamefont {Klimm}}, \bibinfo {author} {\bibfnamefont
  {F.}~\bibnamefont {Simmet}}, \ and\ \bibinfo {author} {\bibfnamefont
  {S.}~\bibnamefont {Horn}},\ }\bibfield  {title} {\enquote {\bibinfo {title}
  {Effect of growth conditions and buffer layers on the metal-insulator
  transition in v2o3 thin films},}\ }\href {\doibase 10.1557/PROC-401-61}
  {\bibfield  {journal} {\bibinfo  {journal} {MRS Online Proceedings Library}\
  }\textbf {\bibinfo {volume} {401}},\ \bibinfo {pages} {61--66} (\bibinfo
  {year} {1995})}\BibitemShut {NoStop}%
\bibitem [{\citenamefont {Walls}\ \emph {et~al.}(2022)\citenamefont {Walls},
  \citenamefont {Murtagh}, \citenamefont {Bozhko}, \citenamefont {Ionov},
  \citenamefont {Mazilkin}, \citenamefont {Mullarkey}, \citenamefont
  {Zhussupbekova}, \citenamefont {Shulyatev}, \citenamefont {Zhussupbekov},
  \citenamefont {Andreev}, \citenamefont {Tabachkova},\ and\ \citenamefont
  {Shvets}}]{Ionov}%
  \BibitemOpen
  \bibfield  {author} {\bibinfo {author} {\bibfnamefont {B.}~\bibnamefont
  {Walls}}, \bibinfo {author} {\bibfnamefont {O.}~\bibnamefont {Murtagh}},
  \bibinfo {author} {\bibfnamefont {S.~I.}\ \bibnamefont {Bozhko}}, \bibinfo
  {author} {\bibfnamefont {A.}~\bibnamefont {Ionov}}, \bibinfo {author}
  {\bibfnamefont {A.~A.}\ \bibnamefont {Mazilkin}}, \bibinfo {author}
  {\bibfnamefont {D.}~\bibnamefont {Mullarkey}}, \bibinfo {author}
  {\bibfnamefont {A.}~\bibnamefont {Zhussupbekova}}, \bibinfo {author}
  {\bibfnamefont {D.~A.}\ \bibnamefont {Shulyatev}}, \bibinfo {author}
  {\bibfnamefont {K.}~\bibnamefont {Zhussupbekov}}, \bibinfo {author}
  {\bibfnamefont {N.}~\bibnamefont {Andreev}}, \bibinfo {author} {\bibfnamefont
  {N.}~\bibnamefont {Tabachkova}}, \ and\ \bibinfo {author} {\bibfnamefont
  {I.~V.}\ \bibnamefont {Shvets}},\ }\bibfield  {title} {\enquote {\bibinfo
  {title} {Vox phase mixture of reduced single crystalline v2o5: Vo2 resistive
  switching},}\ }\href {\doibase 10.3390/ma15217652} {\bibfield  {journal}
  {\bibinfo  {journal} {Materials}\ }\textbf {\bibinfo {volume} {15}} (\bibinfo
  {year} {2022}),\ 10.3390/ma15217652}\BibitemShut {NoStop}%
\bibitem [{\citenamefont {Mazur}\ \emph {et~al.}(2022)\citenamefont {Mazur},
  \citenamefont {Lubańska}, \citenamefont {Domaradzki},\ and\ \citenamefont
  {Wojcieszak}}]{Tendency}%
  \BibitemOpen
  \bibfield  {author} {\bibinfo {author} {\bibfnamefont {M.}~\bibnamefont
  {Mazur}}, \bibinfo {author} {\bibfnamefont {A.}~\bibnamefont {Lubańska}},
  \bibinfo {author} {\bibfnamefont {J.}~\bibnamefont {Domaradzki}}, \ and\
  \bibinfo {author} {\bibfnamefont {D.}~\bibnamefont {Wojcieszak}},\ }\bibfield
   {title} {\enquote {\bibinfo {title} {Complex research on amorphous vanadium
  oxide thin films deposited by gas impulse magnetron sputtering},}\ }\href
  {\doibase 10.3390/app12188966} {\bibfield  {journal} {\bibinfo  {journal}
  {Applied Sciences}\ }\textbf {\bibinfo {volume} {12}} (\bibinfo {year}
  {2022}),\ 10.3390/app12188966}\BibitemShut {NoStop}%
\bibitem [{\citenamefont {{Van Bilzen}}\ \emph {et~al.}(2015)\citenamefont
  {{Van Bilzen}}, \citenamefont {Homm}, \citenamefont {Dillemans},
  \citenamefont {Su}, \citenamefont {Menghini}, \citenamefont {Sousa},
  \citenamefont {Marchiori}, \citenamefont {Zhang}, \citenamefont {Seo},\ and\
  \citenamefont {Locquet}}]{VANBILZEN2015143}%
  \BibitemOpen
  \bibfield  {author} {\bibinfo {author} {\bibfnamefont {B.}~\bibnamefont {{Van
  Bilzen}}}, \bibinfo {author} {\bibfnamefont {P.}~\bibnamefont {Homm}},
  \bibinfo {author} {\bibfnamefont {L.}~\bibnamefont {Dillemans}}, \bibinfo
  {author} {\bibfnamefont {C.-Y.}\ \bibnamefont {Su}}, \bibinfo {author}
  {\bibfnamefont {M.}~\bibnamefont {Menghini}}, \bibinfo {author}
  {\bibfnamefont {M.}~\bibnamefont {Sousa}}, \bibinfo {author} {\bibfnamefont
  {C.}~\bibnamefont {Marchiori}}, \bibinfo {author} {\bibfnamefont
  {L.}~\bibnamefont {Zhang}}, \bibinfo {author} {\bibfnamefont {J.~W.}\
  \bibnamefont {Seo}}, \ and\ \bibinfo {author} {\bibfnamefont {J.-P.}\
  \bibnamefont {Locquet}},\ }\bibfield  {title} {\enquote {\bibinfo {title}
  {Production of vo2 thin films through post-deposition annealing of v2o3 and
  vox films},}\ }\href {\doibase https://doi.org/10.1016/j.tsf.2015.08.036}
  {\bibfield  {journal} {\bibinfo  {journal} {Thin Solid Films}\ }\textbf
  {\bibinfo {volume} {591}},\ \bibinfo {pages} {143--148} (\bibinfo {year}
  {2015})}\BibitemShut {NoStop}%
\bibitem [{\citenamefont {Asmis}\ and\ \citenamefont {Sauer}(2008)}]{clusters}%
  \BibitemOpen
  \bibfield  {author} {\bibinfo {author} {\bibfnamefont {K.}~\bibnamefont
  {Asmis}}\ and\ \bibinfo {author} {\bibfnamefont {J.}~\bibnamefont {Sauer}},\
  }\bibfield  {title} {\enquote {\bibinfo {title} {Erratum: Mass-selective
  vibrational spectroscopy of vanadium oxide cluster ions (mass spectrometry
  reviews (2007) 26 (542-562))},}\ }\href {\doibase 10.1002/mas.20163} {\
  \textbf {\bibinfo {volume} {27}} (\bibinfo {year} {2008}),\
  10.1002/mas.20163}\BibitemShut {NoStop}%
\bibitem [{\citenamefont {Vazquez-Mena}\ \emph {et~al.}(2015)\citenamefont
  {Vazquez-Mena}, \citenamefont {Gross}, \citenamefont {Shenqi}, \citenamefont
  {Villanueva},\ and\ \citenamefont {Brugger}}]{stencil}%
  \BibitemOpen
  \bibfield  {author} {\bibinfo {author} {\bibfnamefont {O.}~\bibnamefont
  {Vazquez-Mena}}, \bibinfo {author} {\bibfnamefont {L.}~\bibnamefont {Gross}},
  \bibinfo {author} {\bibfnamefont {X.}~\bibnamefont {Shenqi}}, \bibinfo
  {author} {\bibfnamefont {L.~G.}\ \bibnamefont {Villanueva}}, \ and\ \bibinfo
  {author} {\bibfnamefont {J.}~\bibnamefont {Brugger}},\ }\bibfield  {title}
  {\enquote {\bibinfo {title} {Resistless nanofabrication by stencil
  lithography: A review},}\ }\href {\doibase 10.1016/j.mee.2014.08.003}
  {\bibfield  {journal} {\bibinfo  {journal} {Microelectronic Engineering}\
  }\textbf {\bibinfo {volume} {132}},\ \bibinfo {pages} {236--254} (\bibinfo
  {year} {2015})}\BibitemShut {NoStop}%
\end{thebibliography}%

\end{document}